# Mobility operator service capacity sharing contract design to risk-pool against network disruptions

**Theodoros P. Pantelidis[1], Joseph Y. J. Chow[1*], Oded Cats[2]**

[1] Department of Civil & Urban Engineering, New York University, Brooklyn, NY, USA
[2] Department of Transport and Planning, Delft University of Technology, Delft, The Netherlands
[*]Corresponding Author Email: joseph.chow@nyu.edu

## Abstract

We propose a new mechanism to design risk-pooling contracts between operators to facilitate horizontal cooperation to mitigate those costs and improve service resilience during disruptions. We formulate a novel two-stage stochastic multicommodity flow model to determine the cost savings of a coalition under different disruption scenarios and solve it using L-shaped method along with sample average approximation. Computational tests of the L-shaped method against deterministic equivalent method with sample average approximation are conducted for network instances with up to 64 nodes, 10 OD pairs, and 1024 scenarios. The results demonstrate that the solution algorithm only becomes computationally effective for larger size instances (above 128 nodes) and that SAA maintains a close approximation. The proposed model is applied to a regional multi-operator network in the Randstad area of the Netherlands, for four operators, 40 origin-destination pairs, and over 1400 links where disruption data is available. Using the proposed method, we identify stable cost allocations among four operating agencies that could yield a 66% improvement in overall network performance over not having any risk-pooling contract in place. Furthermore, the model allows policymakers to evaluate the sensitivity of any one operator's bargaining power to different network structures and disruption scenario distributions, as we illustrate for the HTM operator in Randstad.

## 1. Introduction

Mobility services face the uncertainty of disruptions that can cause significant costs and delays. These disruptions are caused by non-recurrent events (natural or man-made) that reduce transportation supply and can cause breakdown conditions on other parts of the network. The lack of coordination can lead to significant system performance risks and needs to be addressed beforehand to mitigate them. For example, Hurricane Sandy caused $400 million damage to the U.S. public transit system (Frassinelli, 2012) in which 25% of the total cost was from capital expenses.

Traditionally, rail disruptions were addressed via rolling stock and timetable decisions. However, these measures cannot mitigate the immediate consequences of service degradation that cause large system delays due to the lag between operator decisions and the implementation of recovery actions (Kepaptsoglou and Karlaftis 2009). A potential solution is the re-purposing of surplus fleet from other parts of the system (Pender et al., 2013; Liu, 2022), or even from other operators in a multimodal environment or Mobility-as-a-Service (MaaS) platform (Consilvio et al., 2020). In this manner, risks of service degradation due to disruptions are reduced by having multiple operators pool their resources together.

**Definition 1**. *During a disruption,* ***service capacity sharing*** *occurs when one operator with surplus or reserve vehicles in a fleet ((or other interchangeable resources – generally this precludes sharing rail rolling stock to bus routes) repurposes or reassigns them to a disrupted operator in units of passengers that can be served per hour. Examples include bus bridging (when assigning buses to cover train routes) or deadheading reserve vehicles, but across operators.*

As travel becomes essential in large megacities, the benefits of resource consolidation can provide a much better service with fewer delays and disruptions. Such a transport management system requires operators to agree to a contract in advance, which offers a new design problem for mobility operators.

**Definition 2**. *"****Risk-pooling****" (see Levi et al., 2003), widely used in the insurance industry, involves participants contributing small amounts of resources into a pool that they can leverage on during adverse events, thus mitigating the risks of consequential failures.*

The research question that follows is: how to design such a risk-pooling contract between public mobility operators such that they are incentivized to commit service capacity resources that can be used by operators during a disruption? Without pre-disruption agreements, mobility operators may choose not to share resources post-disaster or, worse still, may even leverage the tragedy to exploit further profits (Hawkins, 2018), although the latter case would not work for highly subsidized public transit services since it would cost more to allocate service capacity to exploit the disruption in that case. We propose a service capacity sharing insurance contract mechanism designed to pool the fleets of transportation operators to hedge against disruptions.

The novel problem addressed in this study is distinctive in that the value of risk pooling coalition agreements within a multicommodity flow problem setting has not been studied before under a stochastic multicommodity flow network disruption framework. A stochastic multicommodity flow model is used to determine the optimal amount of resources that each operator needs to contribute to a pool such that the second stage disruption scenarios allow disrupted operators to draw from that pool. These contributions are presented in units of service capacity (e.g., relocating vehicles from one service route to another, such as having a bus serve a connection between two rail stations), not as capacities of road or infrastructure links. The model is coupled with different cost allocation mechanisms to identify conditions needed to ensure stability of the contract; the mechanisms reflect the bargaining power of each operator based on their network structure and committed fleet resources. We propose solution algorithms using an L-shaped method common to stochastic programming and sample average approximation method (SAA) to reach satisfactory solutions for any given coalition in a mobility market.

Computational tests are conducted with the model and algorithms in different-sized instances. For the larger network instance, we evaluate a four-operator insurance contract for the urban multimodal network in the southern ring (Zuidvleugel) of the Randstad area in Netherlands where real disruption data is available. The multi-operator case study network offers an interesting case because it includes multiple operating agencies and the results provide a characterization of their bargaining power in setting up the contract, should their service capacities be interchangeable.

The rest of the paper is organized as follows. Section 2 provides an overview of the literature on two-stage stochastic network problems and game-theoretic profit allocation methods. Section 3 describes the formulation of the two-stage stochastic programming model and solution algorithm. In Section 4, we provide illustrative examples and discuss potential shortcomings of simple allocation mechanisms to further show how cost allocation mechanisms impact the design. We also provide a computational evaluation of the proposed solution methods. Section 5 demonstrates the use of our contract design model in a real case study using data from the multimodal Randstad Zuidvleugel network. Section 6 discusses the key findings and offers concluding remarks and potential future research directions.

# 2. Literature review

## 2.1 Disaster planning methods

Capacity planning is critical for the robustness of transportation networks that face major disruptions (Chen et al., 1999; Cats and Jenelius, 2015). Mitigation against these disruptions has been studied extensively in the context of pre-disruption relief planning. Solution strategies include retrofitting or allocating reserve capacity units in anticipation of disruptions (Miandoabchi and Farahani, 2011; Wang et al., 2015). Other examples of service capacity sharing include redirecting buses from one line to serve a degraded line (e.g. bus bridging and shuttle planning: Kepaptsoglou and Karlafits 2009; Hu et al., 2016; Jin et al., 2016; Van der Hurk et al., 2016; Zhang and Lo, 2020), taxis (e.g. Zeng et al. 2012; Neves-Silva 2013) or for Mobility-on-Demand fleets to provide coverage during disrupted service (e.g. Tyndall, 2009; Yang and Chen 2019; Fang et al., 2020; Cebecauer et al., 2021). Table 1 presents various methods of service capacity sharing that have been proposed in the literature.

**Table 1**: Service capacity sharing studies to deal with disruptions

| Disrupted System | Fleet type | Study |
|---|---|---|
| urban rail/ subway | taxi/bus | Cebecauer et al. (2021) |
| multimodal | multimodal | Consilvio et al. (2020) |
| tram | taxi/bus | Fang et al. (2020) |
| subway | multimodal | Li and Wang (2020) |
| subway | bus | Zhang and Lo (2020) |
| urban rail/ subway | ride-hail | Yang and Chen (2019) |
| subway | bus | Gu et al. (2018) |
| subway | bicycles | Saberi et al. (2018) |
| subway | bus | Zhang and Lo (2018) |
| subway | bus | Jin et al. (2016) |
| multimodal | bus | Van der Hurk et al. (2016) |
| tram | taxi | Neves-Silva (2013) |
| tram | taxi | Zeng et al. (2012) |
| subway | bus | Kepaptsoglou and Karlafits (2009) |
| multimodal | car-share | Tyndall (2009) |

The multicommodity flow problem was first used to model disaster relief planning in Haghani and Oh (1996) as a large-scale deterministic time-space network, which shares common methodologies with transit disruption planning. The two-stage stochastic programming approach has been used extensively in pre-disaster relief network planning (Barbarosoğlu and Arda, 2004; Liu et al., 2009; Rawls and Turnquist, 2010; Peeta et al., 2010; Noyan, 2012; Hong et al., 2015; Klibi et al., 2018; Elçi and Noyan, 2018). A more comprehensive account of two-stage stochastic problems in disaster relief network planning can be found in Grass and Fischer (2016). Common solution methods used in these problems include the L-shaped method for two-stage stochastic programming (e.g. Liu et al., 2009; Rawls and Turnquist, 2010; Miller-Hooks et al., 2012) as well as Monte Carlo-based sample average approximation of the disruption scenarios (e.g. Chen and Yang, 2004; Peeta et al., 2010; Miller-Hooks et al., 2012; Chow and Regan, 2014).

The underlying research gap in those studies is that they assume a single centralized decision-maker. On the contrary, many systems are operated by multiple co-existing operators (Chow and Sayarshad, 2014) in which a centralized operation cannot be assumed, particularly in multimodal networks (see Rasulkhani and Chow, 2019) or MaaS platforms (see Pantelidis et al., 2020; Ma et al., 2019). There are studies of cascading failures in interdependent systems to measure the effects of one system on another (e.g. traffic-electric interactions: Fotouhi et al., 2017). Zhang et al. (2021) examined a bilevel model that considers cooperative game evaluation at an upper level and an integrated computable general equilibrium in the lower, assuming each link is an independent player in a centralized system. Asadabadi and Miller-Hooks (2020) proposed a bi-level game-theoretic formulation to make port investment decisions in a non-cooperative environment under an exogenous coalition structure as a coopetitive game and hedge against disaster events that could affect ports. Liu and Chow (2022) studied coopetitive games for transit operators to share data with each other but did not consider disruptions. There are no studies that deal with the design of incentives for multiple decentralized urban transport operators to collaborate in anticipation of disruptions.

## 2.2 Cooperative games and allocation mechanisms

”Network flow games” deal with multiple operators that own network links and may form coalitions to facilitate network flows (goods or travelers). Early studies looked at minimum tree problems (Bird 1976, Megiddo 1978), the maximum flow problem in Kalai and Zemel (1982) and later the multicommodity flow problem introduced by Derks and Tijis (1985) to minimize the total transportation cost of commodities with different origins and destinations. Curiel et al. (1989) provided a scheme where operators have “committee control” over network arc capacities. Agarwal and Ergun (2008) adopted a game-theoretic mechanism design framework in which links can be owned by multiple operators and enable capacity-sharing cooperation’s to minimize costs subject to individual rationality constraints. Other game-theoretic models involve players other than the network decision-maker either as an attacker (Jin et al., 2015) or an “evil entity” that represents worst case disasters (Bell et al., 2008), leading to a class of network retrofit models called fortification problems under interdiction (e.g. Church and Scaparra, 2007) or network fortification games (Smith and Lim, 2008). We need cooperative game methods that endogenously deal with coalition formation as part of horizontal collaboration (see Doukidis et al., 2007, Lozano et al. 2013) between multiple operators.

To establish a successful mechanism, the joint benefits of collaborating among the members of the collaboration should be distributed in a stable manner (Özener and Ergun, 2008). There are several approaches in the literature to ensure stability of horizontal collaborations. Determining the allocations obtained for participating in a resource pooling contract is just as important as the cost savings estimation model itself. Depending on the value of these allocations, operators may be incentivized to participate in a risk-pooling contract or choose to abstain. Different allocation rules may result in different payoffs for cooperating operators. Myerson (1980) defines these allocation rules as follows:

**Definition 3 (Myerson, 1980).** *An allocation rule is a function* $X: S \rightarrow \mathbb{R}^{|F|}$ *mapping each coalition structure* $s \in S$*, where* $S$ *is the set of all possible coalitions, onto a payoff allocation:* $X(s) = (X_1(s), X_2(s), \dots, X_{|F|}(s))$. $X_f(s)$ *is the payoff allocated to operator* $f \in F$*, under the coalitional structure* $s \in S$. $V(s)$ *is the characteristic function payoff.*

In the equal-gains allocation scheme, we aim to provide the same satisfaction level to all operators. In the proportional method, the satisfaction of operators is proportional to their pool contribution. The total savings $CS(s)$ are the same as before. The core (Gilles, 1953) is a cost allocation concept that ensures no player in a coalition would break away. For example, any game that is convex has a non-empty core and therefore a stable allocation solution. The unique stable set of the convex game coincides with the core. Another mechanism is the Shapley value (Shapley, 1951). The Shapley value is a merit-based allocation mechanism that determines the value of each players contribution to the coalition. It expresses the core center of gravity in convex games. Other mechanisms include the $\tau$-value proposed by Tijs and Driessen (1986), and the nucleolus Schmeidler (1969). The equal satisfaction allocation may not always belong within the core. Only cooperative game-theoretic methods will always belong in the core (when it is non-empty) and consequently guarantee stability and fairness of a coalition.

The Shapley value has been considered as a payoff allocation mechanism in resource pooling contracts (Reinhardt and Dada, 2005). Lozano et al. (2013) use a minimum flow problem to estimate cost savings for every sub-coalition of operators. These savings are allocated using such allocation methods as Shapley value, nucleolus and $\tau$-value. Kellner and Otto (2012) also use a similar approach in allocating $CO_2$ emissions of different shipments in a road transport route. The Shapley value has not been used in combination with a stable solution from a stochastic multicommodity flow problem before.

### 2.3 Summary of research gaps addressed

Our contract design approach uses a stochastic multicommodity flow problem to model costs in the transportation system for a given coalition and allocates capacities based on savings achieved from horizontal cooperation between operators (service capacity sharing). It shares some commonalities with Lozano et al. (2013). A key difference between our study and the study by Lozano et al. (2013), is that the latter is a purely deterministic model. In that study, a core transportation model is solved multiple times to obtain the potential cost savings solution for every sub-coalition of operators. These savings are allocated using cooperative game theoretic (CGT) allocation methods such as Shapley value, nucleolus and $\tau$- value.

We propose a two-stage stochastic model that captures the stochasticity of capacities that are subject to disruptions, which allows us to relate contract design parameters to disruption scenario distributions. Previous stochastic transportation network studies have not considered the contract design problem within the urban mobility market setting and only worked with a limited number of scenarios for much smaller networks. In summary, our contributions include:

- A new model formulation for quantifying the value of a coalition based on a two-stage stochastic multicommodity flow problem to contribute fleets from each operator to a contract so that any disrupted operator can access during a disruption.
- We propose a solution method to solve the model using an L-shaped method for the stochastic multicommodity flow problem with sample average approximation. Computational testing of the methods is conducted for up to 64 nodes, 10 OD pairs, and 1024 scenarios.
- Implementation in a case study of the Randstad area of the Netherlands for four operators, 80 origin-destination pairs, and over 1400 links where disruption data is available to quantify the bargaining power of each operator.

# 3. Proposed methodology

The assumptions made in this model are listed here for convenience.
**A1**. *Mobility operators are assumed to be public, subsidized operators who would not benefit from profiteering off a disrupted operator.*
**A2**. *Service capacities can only be shared if they are interchangeable between the participating operators.*
**A3**. *An operator may be allowed to replenish its service link capacity but not expand its operation further on borrowed fleet.*

## 3.1 Problem statement

A group of operators $F$ own and operate service links (not infrastructure links) $A_f$, $A = \bigcup_{f \in F} A_f$, in a network $G(N, A)$ that serve passengers corresponding to a set of origin-destination (OD) pairs $S$. Operator fleets are represented as link service capacities in the network and are assumed to be continuously transferable, e.g. service lines with vehicle frequencies that may be reassigned to serve another line, with pre-disruption capacities denoted as $w_a$ for each link $a \in A$. Service capacities need to be interchangeable between operators (otherwise they would not be considering a service capacity sharing contract). For example, in general a train and a bus operator would not enter such an agreement because the trains cannot be shared with the bus operator if one of the latter's route was disrupted even if the reverse can be done via bus bridging (Kepaptsoglou and Karlaftis, 2009). Primary users of this framework would be different road-bound public transport systems (Banister and Mackett, 1990), including both fixed route and semi-flexible transit operators (Yoon et al., 2022).

The network is subject to disruption scenarios $\omega \in I$ where one or more links are disrupted, i.e. their capacities are dropped to zero. $\xi(\omega)$ is a binary vector of realized disruptions: $\xi(\omega) \in \mathbb{Z}_2^{|A|}$, where $\mathbb{Z}_2^{|A|}$ is the set of all possible outcomes and is calculated as the Cartesian product of uncertain link capacities in $A$. If a disruption scenario $\omega \in I$ occurs at link $a \in A_f$, then $w_a \xi_a(\omega) = 0$. The model can be trivially extended to consider intermediate capacity degradations instead (see Chow and Regan, 2014).

The contract design problem for service capacity sharing is to determine the equivalent service capacity that each operator $f \in F$ is willing to contribute to a pool, denoted as $b_f$ and should not exceed their total available fleet's capacity. A contribution of $b_f = 0$ implies the operator $f \in F$ does not participate in the pool. In the event of a disruption scenario $\omega$, an impacted operator $f \in F$ can freely borrow service capacity $e_a^f(\omega)$ from the pool to cover link $a \in A_f$. This borrowed fleet is taken from the realized contributions $g_a(\omega)$ taken from links $a \in A_{f'}, g_a(\omega) \leq b_{f'}$, operated by other operators $f' \in F \backslash f$. An operator may be allowed to replenish its service link capacity but not expand its operation further on borrowed fleet $e_a \leq \xi_a w_a$ since this is a resource-pooling contract. A commitment therefore means that an operator is willing to give up to that amount of service capacity to a system optimizer to allocate resources during a disruption. The service capacity sharing contract design problem is illustrated in Figure 1.

To determine the incentives needed for operators to participate, the service capacity sharing agreement assumes that the savings from the service capacity sharing agreement is transferable to the coalition members. These savings do not express attainable operator profits as a unilateral goal but as total societal gains from improved travel times for all network commuters. Thus, this study is distinct in prioritizing the reduction of networkwide delay and then determining an upper bound for public subsidization.

In that case, stability of a service capacity sharing agreement is determined using one of three alternative cost-sharing mechanisms: Shapley value, nucleolus, or $\tau$-value. An empty set implies an unstable contract for that coalition under that mechanism. For a market of operators $F$, different coalitions may be formed. We denote the set of all possible coalitions for a given set of operators $F$ as $\mathcal{V}(F)$ in which a coalition $V_i \in \mathcal{V}(F)$ may form, where $i$ corresponds to the lexicographical order indexing. For example, a market of 3 operators has 8 different possible coalitions: $\{\emptyset, 1,2,3, \{1,2\}, \{1,3\}, \{2,3\}, \{1,2,3\}\}$. Each $i$ corresponds to an index in this ordered sequence, i.e. $V_4 = \{1,2\}$ and $V_0 = \emptyset$ for $\mathcal{V}(\{1,2,3\})$. $\Phi(V_i)$ and $\mathrm{CS}(V_i)$ are correspondingly the costs and cost-savings of members of coalition $V_i$.

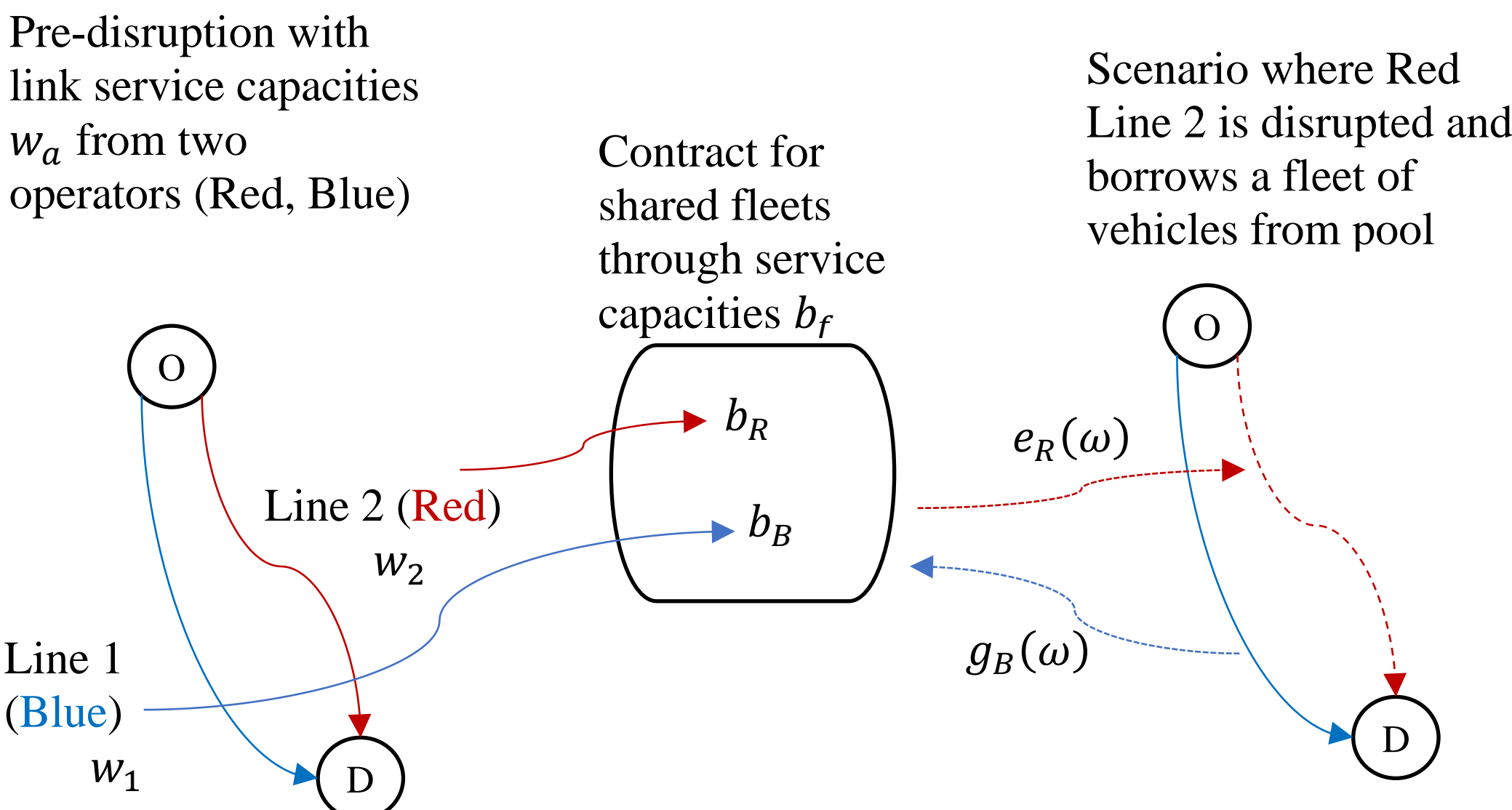

**Figure 1**. Illustration of risk-pooling problem and the role of a service capacity sharing contract.

There are two main challenges in designing the proposed mechanism. The first is to develop a cost estimation function for a given coalition that captures the allocation of service capacity commitments $b_f$ and the value of the coalition. The second is to determine a cost allocation to ensure that the proposed contract is stable. Figure 2 shows how a particular contract design results in cost savings to the coalition.

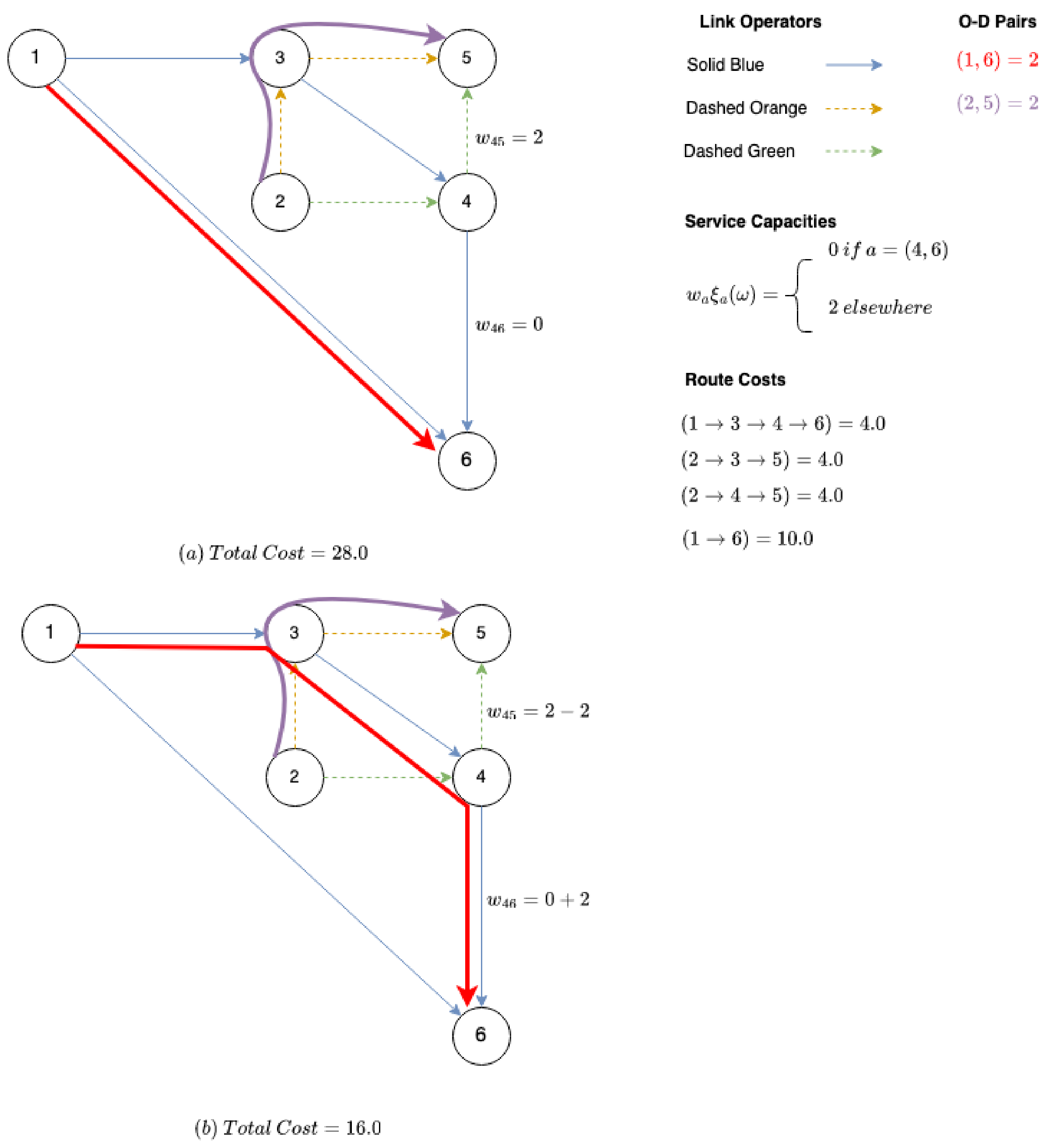

**Figure 2**. Illustration of how service capacity sharing providing cost savings.

Link (4,6) belonging to the blue operator is disrupted in Figure 2(a) and therefore the link's service capacity is set to 0. The total flow costs are estimated to be 28 (in cost units), based on 2 units of flow (1,6) having a cost of 10 each and 2 units (2,5) having a cost of 4 each. If the green operator contributes 2 units of service capacity from link (4,5) to the disrupted link as shown in Figure 2(b), then total network costs can be reduced to 16.

### 3.2 Proposed two-stage stochastic programming cost estimation methodology

We seek an appropriate reward (or cost allocation) to incentivize operators to contribute fleets to improve network costs when a disruption occurs. Figure 3 (top) provides an overview of the mechanism. A service capacity sharing coalition will lead to cost-savings if the network costs are lower than those of standalone operations ($\Phi(V_i) < \Phi(V_0)$). A new two-stage stochastic cost estimation model is proposed to determine the cost of every sub-coalition. A cost-allocation mechanism (Shapley value, nucleolus, etc.) is applied post-hoc to allocate the benefits between the participating operators. While the cost allocation mechanisms are not new, the model formulation for quantifying the value of a coalition is novel. Operators that do not join the coalition are assumed not to form subcoalitions with other non-participating operators $\Phi(\{\emptyset\})$. Figure 3 (bottom) illustrates the types of coalitions that are evaluated.

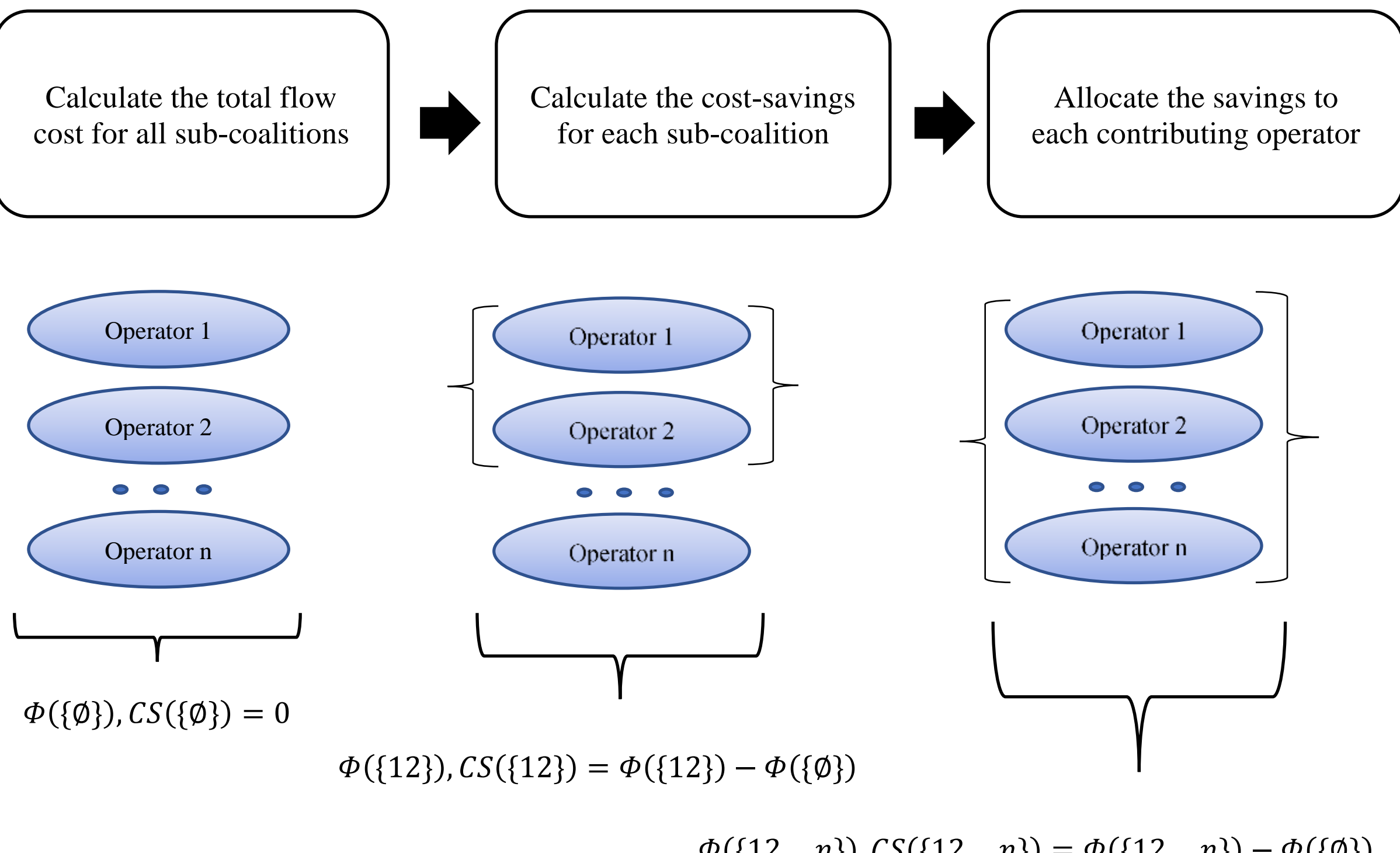

**Figure 3**. Illustration of the mechanism (top) and different coalitions that are evaluated (bottom).

The quantification of the value of a particular coalition is defined as a two-stage stochastic multicommodity flow problem. The goal is to minimize the expected total flow costs given a disruption scenario $\omega \in I$, which can indicate the failure of multiple links (and model correlated network disruptions: see Lo and Tung, 2003; Sumalee and Watling, 2008; Chow and Regan, 2014). The stochastic program is divided into two stages. The first stage decision variables $b^T =$ $(b_1, b_2, .., b_{|F|})$ denote the fleet contributions to the pool, expressed in units of service capacity, by each operator $f \in F$ and are made prior to knowing the outcome of the random disruption scenario. The second stage variables are made after a random disruption event is realized: flows $x_a^s$ on links $a \in A$ for OD $s \in S$, service capacities $g_a, a \in A_f$, made by operator $f$, and excess fleet service capacities $e_{a'}, a' \in A_{f'}$, borrowed by operator $f'$ for link $a'$. The available service capacity is bounded by the first stage decisions.

**Notation**

***Parameters***

$I$: finite set of disruption scenarios

$K$: set of discrete scenarios representing disruption space $I$

$F$: set of operators

$G(N, A)$: network $G$ of nodes $N$ and links $A$ which can be separated into disjoint sets $A_f, f \in F$, and links $A'$ that are not owned by any operator (e.g. alternative modes not considered in market of operators): $A = \cup_{f \in F} A_f \cup A'$

$S$: set of OD pairs

$\xi(\omega) \in \mathbb{Z}_2^{|A|}$: indicator (0,1) vector of length $|A|$ corresponding to link disruptions in scenario $\omega \in I$

$c_a$: travel cost on link $a \in A$

$d^s$: demand amount of O-D pair $s \in S$

$O(s), D(s)$: origin and destination nodes of $s \in S$

$w_a$ : initial service capacity at link $a \in A$

$E_\xi$: expectation across random events $\xi$

$N_i(+)$: set of links inbound to node $i$

$N_i(-)$: set of links outbound from node $i$

***Decision variables***

$x_a^s(\omega)$: flow on link $a \in A_f \cup A'$ for OD pair $s \in S$ in scenario $\omega$

$e_a(\omega)$: service capacity surplus allocated from the pool to link $a \in A_f$ for operator $f \in F$ in scenario $\omega$

$g_a(\omega)$: service capacity deficit contributed to the pool from link $a \in A_f$ by operator $f \in F$ in scenario $\omega$

$b_f$: service capacity contributed by operator $f$ to the pool

Following the above notation, the formulation of the two-stage stochastic programming problem is shown in Eqs. (1) – (14). Eqs. (1) – (3) comprise the first-stage problem and (4) – (14) provide the formulation of the second-stage problem.

$$\Phi(V_i) = \min_b E_\xi Q\left(b^T, \xi(\omega)\right) \tag{1}$$

subject to

$$b_f \geq 0 \qquad \forall f \in V_i \tag{2}$$

$$b_f = 0 \qquad \forall f \in F \backslash V_i \tag{3}$$

where

$$Q\left(b^T, \xi(\omega)\right) \coloneqq \min_x \sum_{s \in S} \sum_{a \in A_f \cup A'} c_a x_a^s \tag{4}$$

Subject to

$$\sum_{a \in N_i(+)} x_a^s - \sum_{a \in N_i(-)} x_a^s = \begin{cases} d^s & if\ i = O(s) \\ -d^s & if\ i = D(s) \\ 0 & \text{otherwise} \end{cases} \qquad \forall i, s \tag{5}$$

$$\textstyle\sum_{s \in S} x_a^s \leq \xi_a w_a + e_a - g_a \qquad \forall a \in A_f, \forall f \in F \tag{6}$$

$$\textstyle\sum_{a \in A_f} e_a \leq \sum_{f' \in F \backslash \{f\}} b_{f'} \qquad \forall f \in F \tag{7}$$

$$\textstyle\sum_{a \in A_f} g_a = b_f \qquad \forall f \in F \tag{8}$$

$$\textstyle\sum_{f \in F} \sum_{a \in A_f} e_a \leq \sum_{f \in F} \sum_{a \in A_f} g_a \tag{9}$$

$$e_a \leq \xi_a w_a \qquad \forall a \in A_f, \forall f \in F \tag{10}$$

$$g_a \leq (1 - \xi_a) w_a \qquad \forall a \in A_f, \forall f \in F \tag{11}$$

$$e_a, g_a = 0 \qquad \forall a \in A_f, f \in F \backslash V_i \tag{12}$$

$$x_a^s \geq 0 \qquad \forall a \in A_f \cup A', f \in F, s \in S \tag{13}$$

$$e_a, g_a \geq 0 \qquad \forall a \in A_f, f \in F \tag{14}$$

Objective function (1) is the minimization of the total expected flow costs over a random disruption event $\xi$. The function $Q$ is a recourse function or expected second-stage value function. $\Phi(V_i)$ represents the cost achieved through the horizontal collaboration of a subset of network operators denoted by $V_i \subseteq F$. Eq. (2) imposes non-negativity constraints on first-stage service capacity commitments $b$. Eq. (3) prevents the non-participating operators from contributing resources to the pool and Eq. (12) prevents them from receiving service capacity units. The second-stage objective (4) is the minimization of total flow costs for a realization of $\xi(\omega)$, where $\omega \in I$.

The service capacity allocations are decided by the contract facilitator (e.g. a public agency) that the coalition members agreed to. Constraint (5) expresses the flow conservation constraints for the second-stage problem. Constraint (6) denotes the link service capacity $w_a\xi_a(\omega)$, where $w_a\xi_a(\omega) = 0$ if $\xi_a(\omega) = 0$, which can be expanded using excess service capacity $e$ or removed from the current link to lend to other links using contributed service capacity $g$. To address the issue of infeasibility of constraint (6), we introduce a set of links $A'$ which represent alternative modes of travel (bike, ride-hailing). The cost $c_a: a \in A'$ is assumed to be higher than the current network shortest path costs. We assume that these links are un-capacitated and therefore not subject to constraints (6-9). Constraints (8) and (9) represent balance conditions that require pool contributions to originate from operators' link service capacities and are bounded by the total pool capacity. These capacity units need not be integer since they may represent service frequencies (see Section 5). Variable $g_a$ represents the deficit on a link $a \in A'$ due to capacity contributed to the pool while $e_a$ is the surplus operators may use from the pool to replenish capacity on link $a \in A'$. Constraint (8) requires that all contributed capacity resources should be committed from network links $a \in A$ that each operator $f \in F$ owns. Constraint (9) requires all surplus to be less or equal to the total amount of capacity contributed to the pool, while constraint (10) allows operators to only be able to replenish their service capacity if there is a disruption and may not use pool resources to expand their network throughput.

Finally, constraints (13) – (14) enforce non-negativity for all second-stage decision variables. The objective value without any pooling minus Eq. (1) indicates the cost savings which translate to the value of the coalition used for determining the stability of the contract agreement.

Under this design, operators are assumed not to be reallocating service capacity from one of their links to another of their own links. This is because we assumed that the capacities represent the steady state pre-disaster capacities which may already capture any existing dynamic resource allocations within the system. A variant mechanism is to allow an operator to allocate service capacity to their own fleet, i.e. $\sum_{a \in A_f} e_a \leq \sum_{f' \in F} b_{f'}$. In that situation, we would need to compute the solo operators as their own coalitions.

The computational complexity of solving a multicommodity flow problem on a directed graph as a linear program has a complexity of $O(n^4)$, where $n$ is the number of nodes. When considering disruption probabilities for each of $m$ links, the model can be expressed equivalently as a deterministic equivalent problem (DEP). The DEP associated with Eqs. (1) – (14) can be formulated when the outcome space $I$ is modeled with a discrete set of scenarios $K$ with probabilities $p_k$ associated with each discrete scenario $k \in K$. Given a number of finite scenarios $k = 1,2, \dots, |K|$ the expected second-stage value function can be expressed as the linear weighted expectation of independent scenario outcomes:

$$E_\xi Q\big(b^T, \xi(\omega)\big) = \sum_{k=1}^{K} p_k c^T x \tag{15}$$

Eq. (15) represents the total expectation over a finite number of scenarios $K$. The second-stage decision variables are expanded into the scenario dimension. The complexity of solving a DEP as a linear program increases to $O(2^m n^4)$, which is not scalable.

### 3.4 Solution method

The proposed model in Eq. (1) – (14) can be solved using stochastic programming methods, but it is not clear which method is most appropriate for this class of problems. We apply and test the use of L-shaped method to decompose the model against the benchmark of simply solving the model as a DEP. The DEP can be solved using a decomposition method commonly known as the L-shaped method because of the block structure of $K$ independent scenarios. This decomposition method was introduced by Van Slyke and Wets (1969) to solve linear stochastic programs and was shown to greatly reduce computational efforts required to generate a solution. An illustration of the block structure in a stochastic program is shown in Figure 4, where $A$ is the first stage constraints, and the second stage constraints can be divided into a scenario dependent portion $T_k, k \in K$, and an independent portion $W$. Matrix $A$ is the set of first-stage constraints in Eq. (2) – (3). Matrix $W$ corresponds to second-stage variable coefficients of Eq. (4) – (14) pertaining to $x, e, g$ while $T_k$ is the first-stage variable coefficients of Eqs. (4) – (14) corresponding to $b$ (Eq. (7) – (8)).

Sample average approximation method (SAA) is another method that can be applied. We consider Monte Carlo simulation to obtain a sample $L$ of the scenarios $K$, known as sample average approximation method (Shapiro and Philpott, 2007). This can be run in combination with the L-shaped method (e.g. Miller-Hooks et al., 2012) or with DEP. This method works best for cases where the total number of scenarios is very large or even infinite. We can generate a sample: $\hat{\xi}(1), \ldots, \hat{\xi}(L)$ of $L$ replications of random vector $\xi$. The SAA does not require scenario enumeration and is appropriate for large network instances.

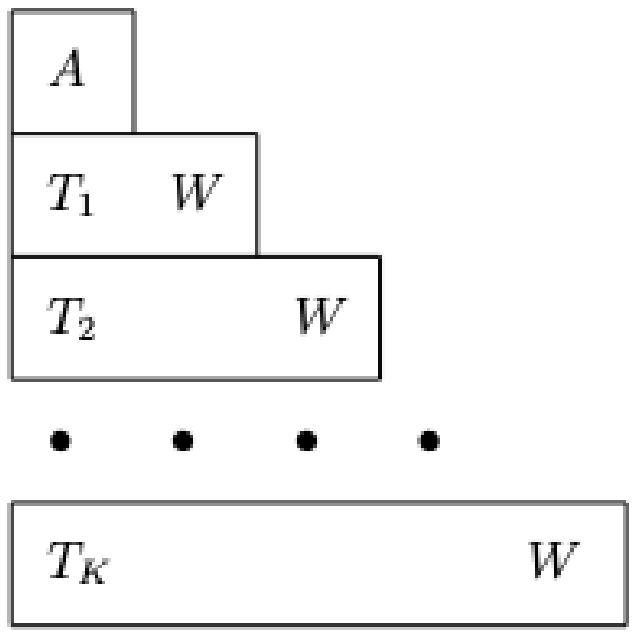

**Figure 4**. Block structure of the L-shaped method. (source: Hoppe, 2007)

The overall solution method is presented as Algorithm 1.

**Algorithm 1: SAA-based L-shaped decomposition**

Inputs: $I, K, F, G, S, V_i$

1. Simulate a set $L \subset K$ of scenarios (see Shapiro and Philpott, 2007)
2. Decompose Eq. (1) – (14) into components for L-shaped method for the scenario set $L$: $A$ is the set of first-stage constraints in Eq. (2) – (3), $W$ is the matrix of second-stage variable coefficients of Eq. (4) – (14) pertaining to $x, e, g$, $T_k$ is the first-stage variable coefficients of Eqs. (4) – (14) corresponding to $b$ (Eq. (7) – (8))
3. Solve Eq. (1) – (14) using L-shaped method (Van Slyke and Wets, 1969)

Outputs: $\Phi, b$

Algorithm 1 is used iteratively to generate the cost estimates for each operator sub-coalition, as shown in Figure 3. In each iteration, we solve Eq. (1) – (14) since the set of participating operators $f' \in F \backslash \{f\}$ in Eq. (7) will be different. This algorithm uses the L-Shaped method to generate a set of scenarios $L$ via Monte Carlo simulation on link failure probabilities to reduce the size of the problem $L \subset K$.

### 3.5 Cost allocation mechanisms used

After determining the commitments of the operators, the stability of the coalition is determined using one of several alternative cost allocation mechanisms reviewed in Section 2. The Shapley formula in Eq. (16) asserts that each player's total gain from a coalition structure is a weighted average of his contributions to all players in smaller coalition structures:

$$Sh_f = \sum_{V \subset F} \frac{(V-1)!\,(F-V)!}{V!} [\Phi(V) - \Phi(V-f)] \qquad (16)$$

The nucleolus also lies in the core if that exists (Schmeidler, 1969). The nucleolus is calculated by finding the imputations that minimize the maximum dissatisfaction. The excess of Eq. (17) represents the dissatisfaction for the coalition V:

$$e(X,V) = \Phi(V_i) - \sum_{j \in V_i} X_j \qquad (17)$$

Also, the core of a game is expressed as a set of imputations: $\sum_{i \in s} X_i \geq \Phi(V)$ . The $\tau$-value is expressed as the unique efficient payoff on interval $[m(V), M(V)]$ (Tijs, 2003). $M(V)$ is the marginal contribution allocation vector, where each element $M_j(V_i)$ expresses the marginal contribution of player $j$ to coalition $V_i$ as shown in Eq. (18).

$$M_j(V_i) = \Phi(F) - \Phi(\mathrm{F}/\{\mathrm{j}\}) \qquad (18)$$

This vector is also called the set of Utopia payoffs. The concept of the minimum rights vector is quite similar to the nucleolus, but instead of minimizing dissatisfaction, $m_j(V_i)$ represents the least-amount that player $j$ can ask in coalition $V_i$. The minimum rights vector $m_j(V_i)$ is shown in Eq. (19).

$$m_j(V_i) = \max_i \left( \Phi(V_i) - \sum_{j \in V_i \backslash \{j\}} M_j(V_i) \right) \qquad (19)$$

## 4. Model verification tests

We demonstrate model feasibility and performance using an illustrative example. We investigate the conditions under which operators will participate in a contract and discuss some key CGT concepts that provide meaningful insights. We then report the results of several cost-allocation methods and discuss some interesting findings.

### 4.1. Illustrative example

Consider the following network instance. Table 2 reports the O-D demand patterns for the illustrative example shown in Figure 5 and Table 3 lists the network parameters.

Let us now consider disruptive events that can potentially disrupt service on links $(1,2)$ and $(2,3)$. Figure 5 shows the probabilities of disruption assumed on the links with an 80% chance of failing completely, i.e. full link closure. There are four distinct failure scenarios $(2^2)$ for this example, computed as the cartesian production of link failures. These risk profiles may correspond to link-specific disruption events such as signal failure, demonstrations, traffic accidents or a tree falling. These independent probabilities are then used to calculate the scenario tree that enumerates all possible outcomes as described in Birge and Louveaux (2011).

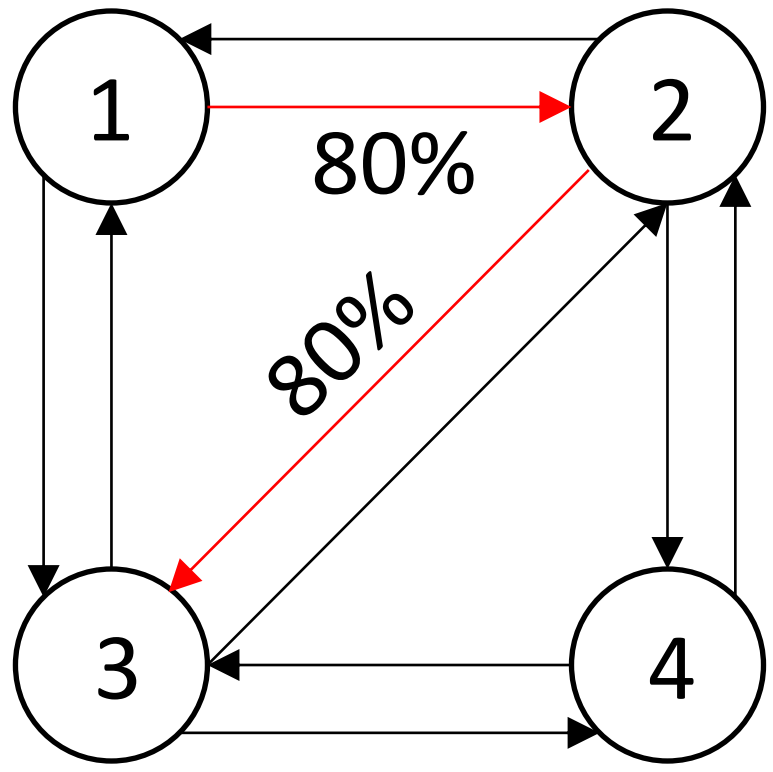

**Figure 5**. Illustrative instance.

**Table 2**. O-D demand

| O-D | Demand |
|---|---|
| (1,2) | 60 |
| (2,3) | 5 |
| (1,4) | 10 |
| (2,4) | 10 |

**Table 3.** Network characteristics

| From node | To node | Travel cost | Service capacity $[f_1, f_2, f_3]$ | Failure prob. $[f_1, f_2, f_3]$ |
|---|---|---|---|---|
| 1 | 2 | 2 | [60, 0, 0] | [**0.8**, 0, 0] |
| 2 | 1 | 2 | [0, 0, 0] | [0, 0, 0] |
| 1 | 3 | 7 | [60, 0, 0] | [0, 0, 0] |
| 3 | 1 | 7 | [0, 0, 0] | [0, 0, 0] |
| 1 | 4 | 10 | [10, 0, 0] | [0, 0, 0] |
| 4 | 1 | 10 | [0, 0, 0] | [0, 0, 0] |
| 2 | 3 | 3 | [0, 5, 0] | [0, **0.8**, 0] |
| 3 | 2 | 3 | [60, 0, 0] | [0, 0, 0] |
| 2 | 4 | 4 | [0, 0, 0] | [0, 0, 0] |
| 4 | 2 | 4 | [0, 5, 10] | [0, 0, 0] |
| 3 | 4 | 3 | [0, 5, 15] | [0, 0, 0] |
| 4 | 3 | 3 | [0,0,20] | [0,0,0] |

#### *4.1.1. L-shaped method illustration*

The L-shaped solution method in Algorithm 1 generates a total of 13 optimality cuts and 8 feasibility cuts before it reaches the optimal objective value, $E_{\xi}Q\left(b^T, \xi(\omega)\right) = 499$. Table 4 illustrates the results of the service capacity sharing model obtained solving the L-shaped method. The scenario columns correspond to the second-stage problem variables while the objective value and pooled service capacity are results of the first stage problem. The results were verified by solving the deterministic equivalent program (DEP) and found to be the same.

**Table 4**. Results of service capacity sharing model for small instance

| Flow Variables X[link, O-D pair, operator] | Scenario #1 | #2 | #3 | #4 | Pooled service capacity $(\boldsymbol{b}(\mathbf{1}), \boldsymbol{b}(\mathbf{2}), \boldsymbol{b}(\mathbf{3}))$ | Objective value |
|---|---|---|---|---|---|---|
| **X[(1,2), (1,2), 1]** | 60 | 60 | 25 | 25 | $(5, 5, 20)$ | 499 |
| **X[(2,3), (2,3), 2]** | 5 | 5 | 5 | 5 | | |
| **X[(1,4), (1,4), 3]** | 10 | 10 | 10 | 10 | | |
| **X[(2,4), (2,4), 2]** | 5 | 5 | 5 | 5 | | |
| **X[(2,4), (2,4), 3]** | 5 | 5 | 5 | 5 | | |
| **X[(1,3), (1,2), 2]** | 0 | 0 | 35 | 35 | | |
| **X[(3,2), (1,2), 2]** | 0 | 0 | 35 | 35 | | |

The results in Table 4 suggest that service capacity contributions of $b = (5, 5, 20)$ are needed to maintain optimal flows. Operator 3 contributes the largest amount, 20 units of service capacity to the pool while operators 1 and 2 contribute only 5 units. Even if an operator does not contribute any resources, the total value of the objective may still be reduced by being able to use the pooled resources and thus provide more service capacity to its users. This raises the question of what an appropriate compensation scheme for each operator would be to participate in resource pooling.

#### *4.1.2. Operator savings allocations*

We now employ the procedure described in Figure 3 to calculate the cost-savings for each coalition in our illustrative example. The results are given in Table 5. We use the measure of synergy to evaluate the importance of a coalition. The synergy of a coalition $V_i$ is given by the ratio of savings divided by the total costs shown in Eq. (20).

$$Synergy(V_i) = \frac{CS(V_i)}{\Phi(V_i)} \tag{20}$$

The measure of synergy is very important in understanding the effectiveness of each coalition. It can be a very useful tool for government agencies when deciding which transportation providers to invite to take part in the mobility contracts.

**Table 5**. Coalition results

| Coalition $V_i$ | $\Phi(V_i)$ | $CS(V_i)$ | $Synergy(V_i)$ |
|---|---|---|---|
| $\{\emptyset\}$ | 675 | 0 | 0.00 |
| $\{1\}$ | 675 | 0 | 0.00 |
| $\{2\}$ | 675 | 0 | 0.00 |
| $\{3\}$ | 675 | 0 | 0.00 |
| $\{12\}$ | 595 | 80 | 0.13 |
| $\{13\}$ | 579 | 96 | 0.17 |
| $\{23\}$ | 658 | 17 | 0.03 |
| $\{123\}$ | 499 | 176 | 0.35 |

Table 6 summarizes the results obtained under different allocation rules for the coalition of three operators $\{123\}$. While there is no ubiquitous "appropriate" allocation method, we consider the Shapley value to be an adequate indicator for resource pooling problems. Apart from game-theoretic allocation mechanisms, we consider an equal satisfaction outcome and a proportional-to-contribution reward allocation method. Different allocation rules distribute subsidy amounts differently to operators conforming to the criteria that are set by each allocation rule $X(V_i)$. Computations were performed in TUGlab (Calvo and Rodriguez, 2006) and MatTU Games (Meinhardt, 2020).

**Table 6**. Comparison of different allocation mechanisms

| Allocation rule / Operator | Equal Satisfaction | Proportional contribution | Shapley value | Core center | Nucleolus | $\tau - value$ | Utopia |
|---|---|---|---|---|---|---|---|
| 1 | 58.66 | 29.33 | 79.83 | 79.65 | 79.83 | 79.77 | 160.00 |
| 2 | 58.66 | 29.33 | 48.33 | 48.43 | 48.33 | 48.36 | 97.00 |
| 3 | 58.66 | 117.33 | 47.83 | 47.93 | 47.83 | 47.86 | 96.00 |

*subsidy amounts are given in terms of travel cost savings

The game presented in this section happens to be non-convex (supermodularity condition is violated) but superadditive, and thus all CGT methods listed in Table 6 are core allocations that maintain stability of the service capacity sharing agreement. However, if the operators agree to a cost allocation of the benefits based on the amount of resources that each operator contributes (proportional contribution), then the allocation vector would not belong in the core. The proportional method is especially unreliable since cost-savings model solutions may not be unique.

Figure 6 depicts the core of the game. Every payoff vector that lies in the core (dark polygon) is considered a fair and stable solution.

### 4.2 Performance evaluation

We compare the performance of the solution method (Algorithm 1 with L-shaped method) with benchmark methods: DEP with enumerated scenarios, and SAA-DEP. The computational results are obtained using Gurobi 9.5.1 in Python 3.8.9 on a 13'' MacBook Air laptop i-7 laptop with 8GB RAM 1600MHZ-DDR3 and OS Monterey 12.3.1. These algorithms are coded from scratch in Python for consistency purposes (e.g., SAA runtimes benchmarked against the L-shaped method). To compare the performance of these solution methods, we generate 6 random grid network instances. Table 7 provides a summary of parameters used to create these test instances. The instances can be found in Pantelidis (2020).

Link operators and vulnerable links are assigned randomly, and their respective values are drawn from formulas listed in Table 7. For each O-D pair an alternative path is generated to ensure that flow constraints are met. The results are summarized in Table 8. The sizes of these examples far exceed the largest examples in the literature (e.g. Miller-Hooks et al., 2012).

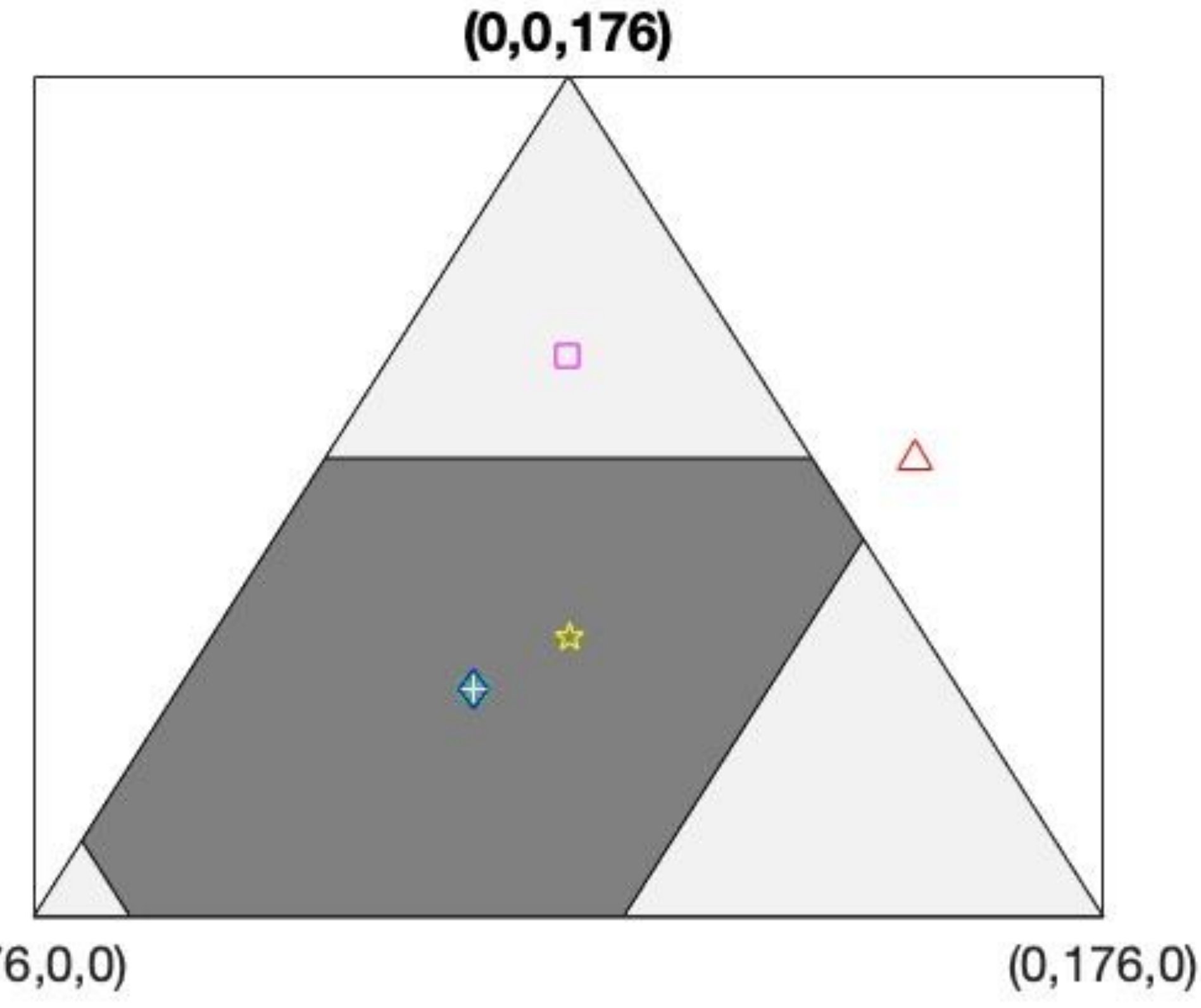

**Figure 6**. The core of the game (dark grey area) and allocation methods of table 5: the core center (white cross), Nucleolus (green circle), $\tau$-value (blue diamond), Shapley value (cyan asterisk), equal satisfaction (yellow star), Utopia payoffs (red triangle) and proportional contribution (magenta square).

**Table 7**. Random network parameters

| Parameter | Value |
|---|---|
| # of links | $2(N-\sqrt{N})$ |
| $w_{ij}$ | Random INT$[0, N^2]$ |
| $c_{ij}$ | Random INT$[0, 100]$ |
| # of scenarios | $2^{(\sqrt{N}+2)}$ |
| # of O-D pairs | $\sqrt{N}+2$ |
| $d^s$ | Random INT$[0, N^4]$ |
| $p_{ij}$ | Random [0, 1] |

**Table 8**. Summary of test results

| | Runtime (seconds) | | | | Absolute Optimality Gap (%) | |
|---|---|---|---|---|---|---|
| # of scenarios [Exact, SAA] | DEP | L-Shaped | SAA (DEP) [avg, std]* | SAA (L-Shaped) [avg, std]* | SAA (DEP) | SAA (L-Shaped) |
| [16, 32] | 0.16 | 1.03 | [0.37, 0.04] | [1.85, 0.24] | 3.78 | 3.78 |
| [32, 32] | 1.19 | 2.72 | [1.11, 0,18] | [2.26, 0.09] | 0 | 0 |
| [64, 32] | 7.50 | 15.18 | [2.60, 0.05] | [7.45, 0.38] | 4.89 | 4.89 |
| [128 ,32] | 99.16 | 114.82 | [5.92, 0.21] | [19.91, 2.85] | 0 | 0 |
| [256, 64] | 1688.58 | 387.33 | [55.57, 5.08] | [100.03, 12.53] | 1.22 | 1.22 |
| [512, 128] | 16573.81 | 1323.31 | [748.94, 11.04] | [353.99, 12.88] | 0.59 | 0.59 |
| [1024, 256] | 142489.49 | 5801.12 | [8133.04, 63.41] | [1371.36, 176.80] | 0.08 | 0.08 |

* SAA results display are reported average and standard deviation of 10 runs

The single-cut L-shaped algorithm is much more efficient than the DEP once the number of scenarios grows sufficiently large. More scenarios cause a linear increase in the L-shaped runtime but an exponential increase in the DEP runtime.

Overall, Gurobi solution times were very similar using different solution methods (simplex, dual simplex and Barrier method). The technological advances found in modern solvers (sparse matrices, multi-threaded computing and pre-solve methods) have improved the solution times of large linear programs. SAA proves to be adequate for such problems both in solution quality (less than 0.1% optimality gap in some cases) and runtime.

Based on the results reported in Table 8, SAA can be solved quite efficiently alongside DEP when the sample size is sufficiently small. There exists a threshold over which large sample sizes can be solved much more computationally efficiently using SAA with the L-shaped method than with DEP.

## 5. Randstad network case study

We apply our modelling approach to design a contract between the operators of the public transport network of part of the Randstad Zuidvleugel region in the Netherlands shown in Figure 7. The Randstad Zuidvleugel is the southern ring of the Randstad and was used as a case study in Cats et al. (2016) to identify critical links and study disruption effects. The high demand intensity and the large number of multimodal routes imply that disruption costs will be significant for both travelers and operators. Yearly passenger disruption costs resulting from disruptions on one single light rail link in the case study network can exceed €900,000 (Cats et al., 2016). For this reason, we identify a need for operators to hedge the disruption risks and insure the most critical routes. For the Randstad network, a large dataset is available for the period between January 2011-August 2013. In practice, such a multimodal system would not be able to benefit from the contract design model because the rail-bound services cannot be assigned to substitute disrupted road-bound routes; however, for the sake of illustrating the methodology using real disruption data we relaxed this practical requirement here. As a hypothetical case study, the results can provide a potential justification for why the transit operators in that region may want to adopt interoperable vehicular technologies.

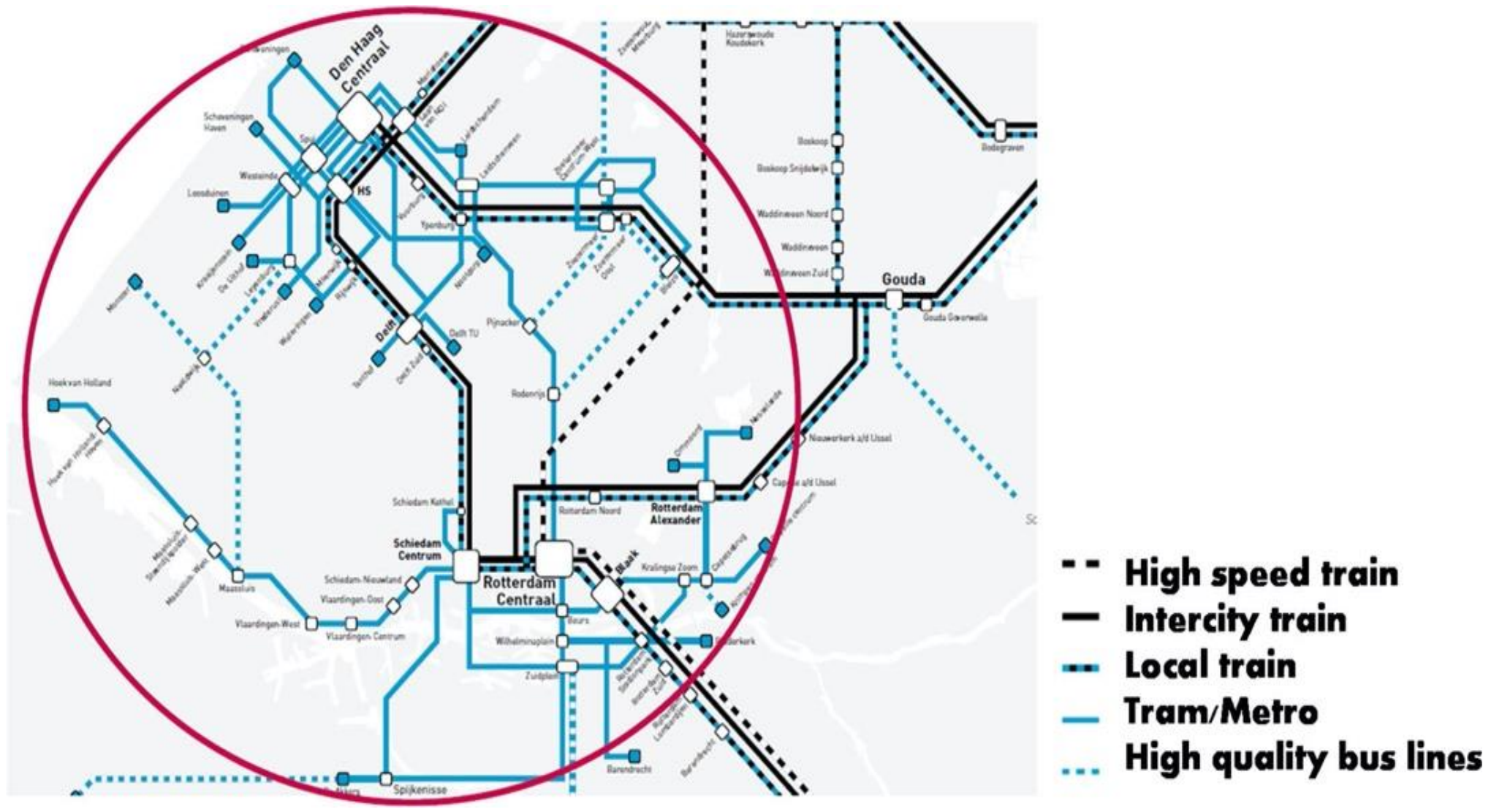

**Figure 7**. The Randstad network. Source: (Cats et al., 2016)

### 5.1. Network operators

During the analysis period, the case study network consists of six different types of services that are ultimately owned by four transportation entities: Dutch Railways (NS), HTM, RET and Connexxion (now EBS). Table 9 presents a summary of transportation services provided in the case study network and Figure 8 presents the GIS representation of the 4 operators' networks generated from GTFS data. Note that the values are not route-level but rather represent summations of link level values.

**Table 9.** Randstad network parameters

| Operators / Parameters | Dutch Railways | | HTM | | RET | Connexion |
|---|---|---|---|---|---|---|
| | **NS-I** | **NS-L** | **HTM** | **HTMBuzz** | | |
| **Sum of all link service capacities** $(\mathbf{1000} \times \boldsymbol{passengers/\ hour})$ | 475 | 250 | 2145 | 464 | 13 | 4 |
| **# of Failures (7-9AM)** $[\boldsymbol{average}, \boldsymbol{std}]$ | [3.75,1.76] | [4.83,2.03] | [63.64,7.08] | [18.82,4.23] | [4.74,1.31] | [2.26,1.14] |
| **Average link frequency (trips per hour)** | 1.60 | 0.45 | **0.01** | **0.01** | 1.22 | 0.32 |
| **Demand** $(\mathbf{1000} \times \boldsymbol{passengers} / \boldsymbol{hour})$ | 13.39 | 19.02 | 4.50 | 0.29 | 4.42 | 0.78 |
| **Total network length (km)** | 131.76 | 223.04 | 298.82 | 196.43 | 72.44 | 32 |

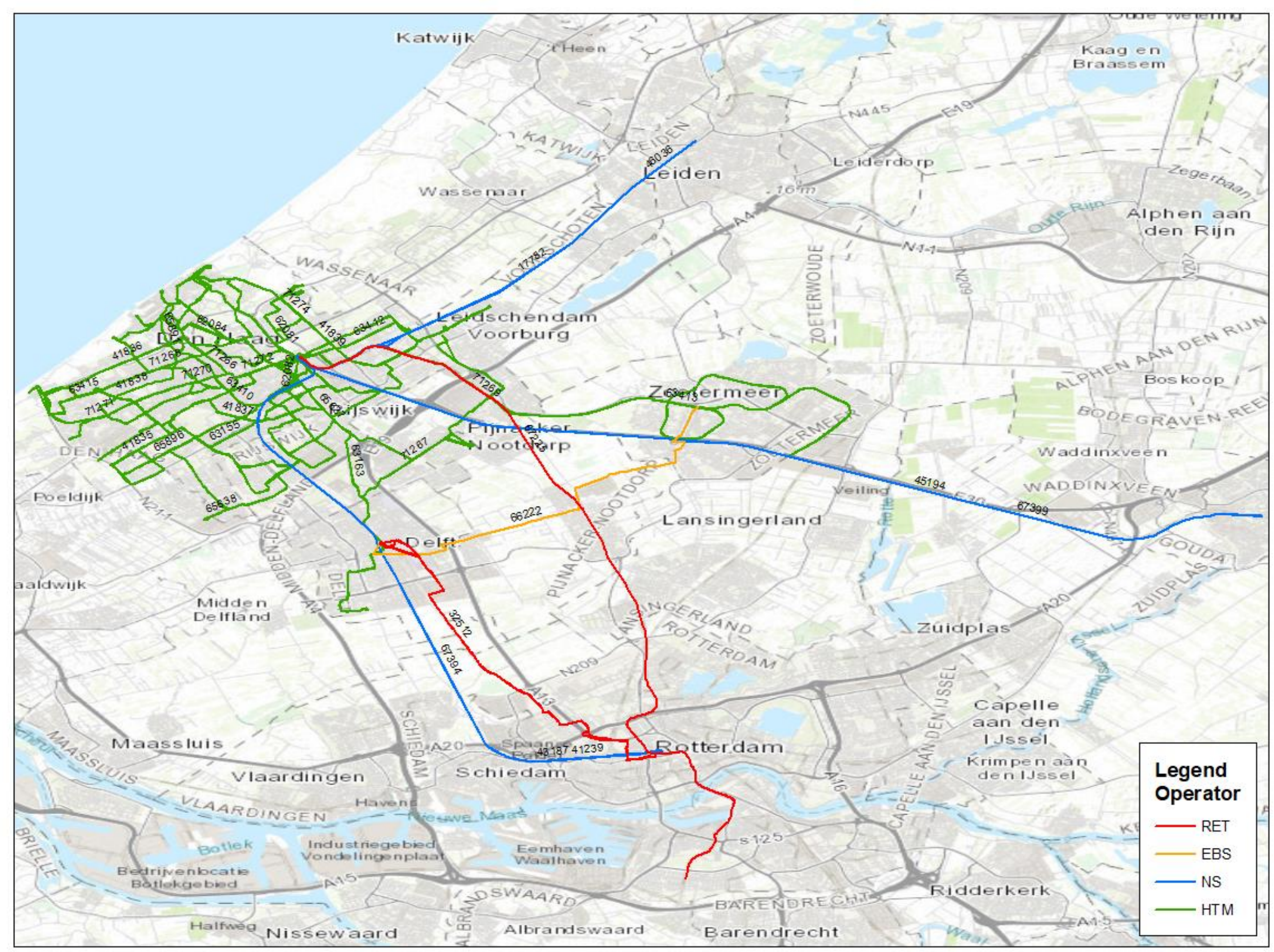

**Figure 8.** GIS map of the four operators for the case study from GTFS data.

The network shown in Figure 8 includes two train services (NS-I, NS-L) that belong to the national railway company NS (Dutch Railways), the Hague-based HTM urban rail (tram and light rail) service that also owns an urban bus fleet that until 2019 operated under the HTM-Buzz flag, the Rotterdam-based RET that runs the metro/light rail service between Hague and Rotterdam and a fleet of buses. Finally, the international company Connexxion owned by Transdev owns a bus fleet that provides service between Delft and Zoetermeer. It is now EBS.

Some of these transportation providers can be considered part of larger transportation agencies in terms of strategic planning and decision making. The intercity train service (NS-I) is operating between Leiden Central, The Hague, Rotterdam and Gouda. NS-L is the local train service. All train services are cordoned at Leiden Central / Gouda / Rotterdam Central. The NS (Dutch Railways) is responsible for operating both the intercity and local services in the case study area and are illustrated in Figure 7 (dashed and solid black lines).

### 5.2. Network characteristics

Within this area there is a high-density public transport network consisting of train, metro, light rail, tram, regional and urban bus services that serve more than 400,000 commuters daily. An origin-destination matrix was approximated by a combination of empirical data (passenger counts, smart card records), transport demand model estimations and for certain routes for which no information was available estimated based on the available capacity and assumed occupancy rates.

Passenger demand and service supply correspond to the 2-hour AM peak period (7-9 a.m.). Link capacity is defined as the product of the crush capacity per public transport vehicle and the number of public transit trips within the AM period. We assume that links are unidirectional and travel costs are calculated according to average running time of link services. The total network capacity is 3,351,000 passengers/hour.

We compute coalition values, pool contributions and subsidies for the four operators present in the case study network. For this experiment we use the 40 highest-volume O-D pairs with a total demand amounting to 223,800 passengers/hour for the entire region. Some of these pairs have unique transit network paths to their destination which may be disrupted leading to infeasible solutions due to violating constraint (5). To account for that, we assume that these travelers can also use alternative modes of transportation such as carpooling, bike, taxi, etc. These alternatives are presented as additional un-capacitated direct-connection links for each network O-D pair. The cost of these links $a \in A'$ will be calculated as: $10 \times (cost\ of\ shortest\ path)$, where the cost of shortest path is the minimum travel path of each pair in the Randstad transit network. Table 9 presents the basic parameters of the network that are used to run the computation experiments.

### 5.3. Disruption data

The network disruption data reflect the disruption exposure by calculating the expected time that a link is exposed to disruptions per time period $(frequency \times duration)$ compared to the total time public transport services are provided on this same link. Based on this premise, failure probabilities are derived from the ratio: $(exposure\ time\ /\ total\ operating\ hours)$. The network disruption data were obtained from Cats et al. (2016). The disruption log information was obtained from different service providers via an API. This data contained information about disruptions for a period of 2.5 years for the train network (NS-I, NS-L), and for a period up to 3 months for the tram and bus networks. Note that disruption data is for databases dating back to 2013. We conduct 10000 randomized Bernoulli trials to estimate the distribution parameters (mean, standard deviation) of the number of disruptions that may occur during the morning peak hours (7-9AM) by using the exposure time over the total operation time ratio. We can see that HTM is more vulnerable to disruptions which can be attributed to higher service frequencies. The demand in Table 8 represents the commuters' willingness to use public transit services by identifying the least-cost route of each O-D pair. Since the network includes more than 1400 links, the number of scenarios is over $2^{1400}$ and hence network disruption effects can only be captured using Algorithm 1 with the SAA approach.

### 5.4. Results and analysis

In this section we present a summary of the computational results using SAA. The algorithm was coded in Python 3.8 using Gurobi 9.5.2. To handle the scale of this experiment, an e2-standard-8 Linux virtual machine was used in Google Cloud with an eight-core CPU and 32GB of RAM. To launch the Gurobi solver in a cloud environment we used a Web License Service (WLS) and executed the script in a containerized environment using Docker.

The service capacity sharing model needs to be run for every sub-coalition, resulting in 4! iterations to obtain the total transportation costs for each coalition combination. We limit the sample size of the problem to $N = 500$ but each iteration is repeated 5 times, so there were $5 \times 4!$ solver runs in total. The objective value average and standard deviation was extracted for each sub-coalition. The total runtime exceeded 72 hours for each run. Generally for planning and provisioning purposes, runtimes in the order of hours are considered acceptable (for example, runs of regional planning models like NYC's BPM or MATSim can take days). What's more, this particular example represents the upper echelon of instance complexity; in practice, transit operators will likely work with coarser segment level networks that would be on the order of low hundreds of links, not thousands as we have here. Significant computational savings can be achieved by means of parallel computing and aggregating solutions from multiple runs where each run has a reduced number of scenarios. Finally, more targeted sampling methods (e.g., Latin Hypercube Sampling) may lead to a more representative set of samples and thus reduce the need for higher number of scenarios.

#### *5.4.1. Coalition savings results for the baseline scenario*

Table 10 provides a summary of the results. Since each iteration is solved multiple times (5 times), we report the average and standard deviation of the objective value. Cost-savings and synergy measures are computed using average values of $V_i$.

**Table 10**. Savings (in million passenger-minutes) and Synergy

| Coalition $V_i$ | $\Phi(V_i)$ [avg , std] | $CS(V_i)$ | $Synergy(V_i)$ |
|---|---|---|---|
| {Ø} | [193.95 , 1.96] | 0.0 | 0.00 |
| {1} | [193.95 , 1.96] | 0.0 | 0.00 |
| {2} | [193.95 , 1.96] | 0.0 | 0.00 |
| {3} | [193.95, 1.96] | 0.0 | 0.00 |
| {4} | [193.95 , 1.96] | 0.0 | 0.00 |
| {12} | [107.19 ,0.05] | 86.76 | 0.81 |
| {13} | [133.24,0.18] | 60.71 | 0.46 |
| {14} | [173.15 , 1.96] | 20.8 | 0.12 |
| {23} | [153.23 ,1.29] | 40.72 | 0.27 |
| {24} | [176.81,0.81] | 17.14 | 0.10 |
| {34} | [171.94 ,0.38] | 22.02 | 0.13 |
| {123} | [75.75 , 0.22] | 118.2 | 1.56 |
| {124} | [99.14 , 0.23] | 94.81 | 0.96 |
| {134} | [118.04 , 0.97] | 75.91 | 0.64 |
| {234} | [142.62 ,1.19] | 51.33 | 0.36 |

| {1234} | [65.66 , 0.01] | 128.29 | 1.95 |
|---|---|---|---|

As shown in Table 9, a service capacity sharing contract can improve network performance by almost 66% ($CS(\{1234\}) = 128.29$, which is 66% of $\Phi(\{\emptyset\}) = 193.95$) which is remarkable. The measure of synergy is indicative of the efficiency of a coalition. This measure can be a very useful tool for decision-making and can assist government agencies in identifying critical operators that can benefit horizontal cooperation schemes significantly.

After identifying these cost-savings coalitions, we report several cost-allocation methods that have desirable properties in terms of fairness and stability in Table 11. These values are reported in million passenger-minutes, which can then be monetarized by choosing an appropriate value of time parameter.

**Table 11.** Cost allocations (in million passenger-minutes)

| **Allocation rule** / **Operator** | Shapley value | Nucleolus | $\tau$-value | Utopia |
|---|---|---|---|---|
| Dutch Railways (NS) | 50.68 | 61.29 | 60.69 | 76.96 |
| HTM | 38.55 | 36.72 | 36.11 | 52.38 |
| RET | 28.72 | 25.22 | 25.25 | 33.48 |
| Connexxion | 10.34 | 5.05 | 6.24 | 10.09 |

Every allocation rule identifies Dutch Railways (NS) as the most critical operator in the network, followed by HTM. The reasoning behind a large cost allocation for NS is due to its large demand. Since many travelers' desires to use the NS network for their itinerary, receiving fleet support from an insurance pool will improve passenger travel times. As shown in Figure 8, HTM offers a highly versatile network with many redundant paths available. RET forms the main connection between Rotterdam and Den Haag. Those could be potential reasons for their bargaining power.

The game is non-convex since supermodularity is violated (e.g. $\Phi(\{123\}) - \Phi(\{12\}) > \Phi(\{1234\}) - \Phi(\{124\})$). The game core is illustrated in Figure 9 along with its four edges. Every subsidy allocation combination that lies inside the core represents a stable outcome. The core provides unique insights in determining the strength of each sub-coalition. The 4-dimensional core can be further broken down in 3-dimensional surfaces. While the grand coalition offers the greatest value, we can also see that coalition {134}, corresponding to a resource pooling contract between HTM, RET and Connexion, has the largest core area surface compared to other 3-operator coalitions. This suggests that they have more space for negotiating cost allocations in an agreement and would then be less prone to perturbations. This sort of analysis can provide important insights into setting strategic goals and designing insurance contracts. The surface of the area of the core shows which players (or operators) can find a common ground more easily than others.

### *5.4.2. Coalition savings results with service capacity reduction*

In this section, we assume a second scenario where NS operates on pre-disruption service capacities reduced by 80% of the original in Table 8. The network parameters are given in Table 12, where the remaining 20% capacities are highlighted in bold. This sensitivity analysis shows how the reduced service capacity impacts NS's bargaining power in the contract design.

Table 13 provides a summary of the results. The savings of the grand coalition are now only 25% of the total flow costs, which is quite small compared to 66% in the base case. The service capacity reduction for NS accounts for 17% of total network service capacity reduction (from 580 to 145) but total costs are increased by 44% (from 193.95 to 347.16 million passenger-minutes) implying that NS contributes many resources to the pool.

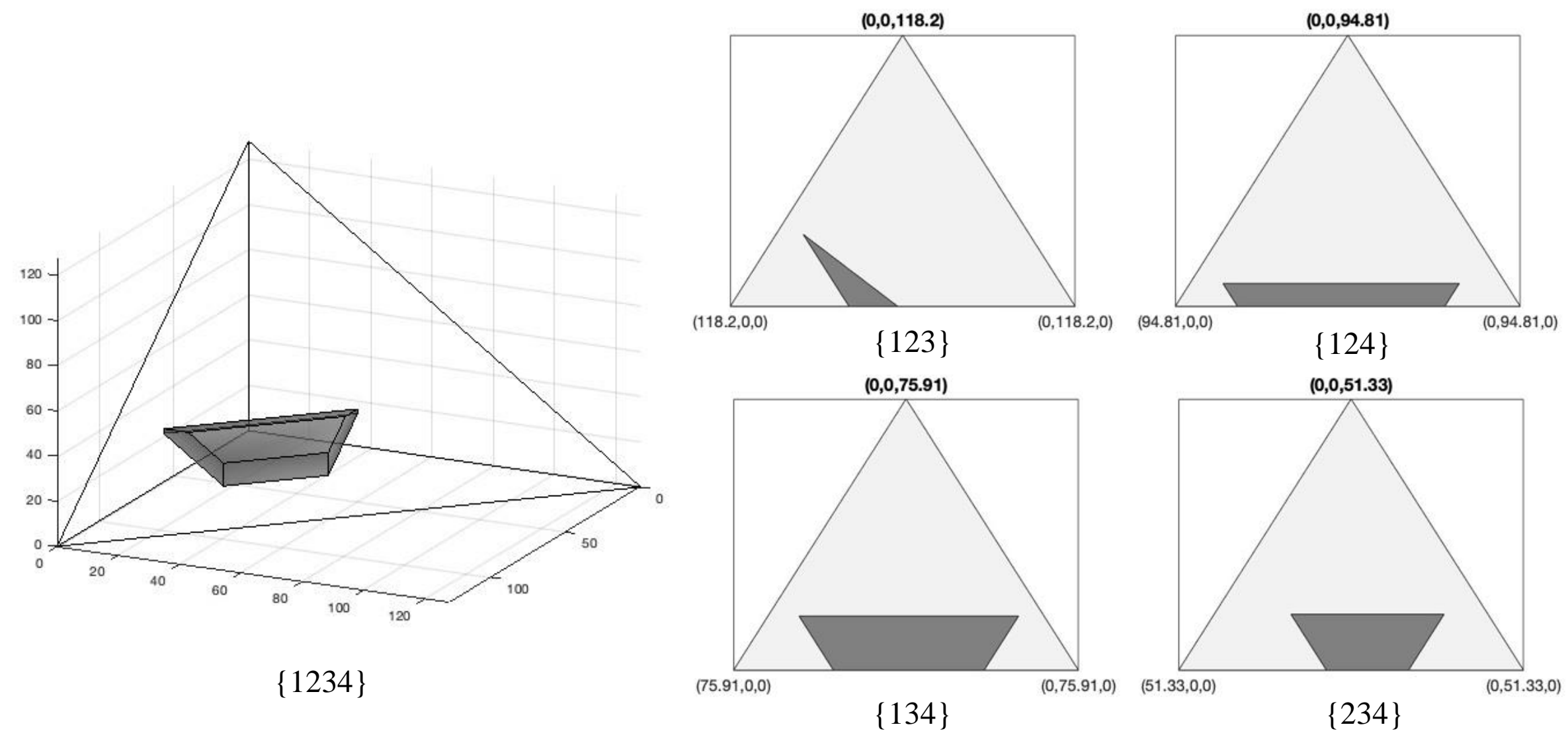

**Figure 9**. Core results for the Randstad operators (baseline).

**Table 12**. Randstad network parameters (under 20% NS service capacity scenario)

| **Operators** / **Parameters** | **Dutch Railways** | | **HTM** | | **RET** | **Connexion** |
|---|---|---|---|---|---|---|
| | **NS-I** | **NS-L** | **HTM** | **HTMBuzz** | | |
| **Sum of all link service capacities** ($1000 \times passengers/\ hour$) | 95 | 50 | 2145 | 464 | 13 | 4 |
| **# of Failures (7-9AM)** [$average, std$] | [3.75,1.76] | [4.83,2.03] | [63.64,7.08] | [18.82,4.23] | [4.74,1.31] | [2.26,1.14] |
| **Average link frequency (1000 trips per hour)** | 1.60 | 0.45 | **0.01** | **0.01** | 1.22 | 0.32 |
| **Demand** ($1000 \times passengers /hour$) | 13.39 | 19.02 | 4.50 | 0.29 | 4.42 | 0.78 |
| **Total network length (km)** | 131.76 | 223.04 | 298.82 | 196.43 | 72.44 | 32 |

**Table 13**. Savings (in million passenger-minutes) and Synergy (under 20% NS service capacity scenario)

| Coalition $V_i$ | $\Phi(V_i)$ **[avg , std $10^{-4}$]** | $CS(V_i)$ | $Synergy(V_i)$ |
|---|---|---|---|
| {∅} | [347.16 ,3.39] | 0.0 | 0.00 |
| {1} | [347.16 ,3.39] | 0.0 | 0.00 |
| {2} | [347.16 ,3.39] | 0.0 | 0.00 |
| {3} | [347.16 ,3.39] | 0.0 | 0.00 |
| {4} | [347.16 ,3.39] | 0.0 | 0.00 |

| {12} | [302.55 ,0.76] | 44.61 | 0.15 |
|---|---|---|---|
| {13} | [302.16 ,0.12] | 45.00 | 0.15 |
| {14} | [329.28 ,0.32] | 17.89 | 0.05 |
| {23} | [306.25, 0.11] | 40.92 | 0.13 |
| {24} | [330.79,1.69] | 16.37 | 0.05 |
| {34} | [326.00, 3.17] | 21.17 | 0.06 |
| {123} | [271.13, 1.32] | 76.03 | 0.28 |
| {124} | [294.98,0.63] | 52.18 | 0.18 |
| {134} | [292.03, 0.52] | 55.14 | 0.19 |
| {234} | [296.76, 0.60] | 50.40 | 0.17 |
| {1234} | [261.80, 0.10] | 85.37 | 0.33 |

From Table 14 we can see that the amount of cost allocation that NS should be getting is significantly lower than before. This means that they would agree to a much lower portion of the savings due to reduced service capacity. For example, based on the Nucleolus allocation vector we can see that the subsidy amount is reduced by 53% (from 61.29 to 29.07 million passenger-minutes).

**Table 14**. Cost allocations under the 20% NS service capacity setting (in million passenger-minutes)

| **Operator** \ **Allocation rule** | Shapley value | Nucleolus | $\tau - value$ | Utopia |
|---|---|---|---|---|
| Dutch Railways (NS) | 26.44 | 29.07 | 28.56 | 34.97 |
| HTM | 23.93 | 24.33 | 23.83 | 30.23 |
| RET | 25.78 | 27.29 | 26.55 | 33.19 |
| Connexxion | 9.22 | 4.69 | 6.43 | 9.34 |

The core also confirms that NS has reduced bargaining power. For example, coalition {124} shown in Figure 10 is clearly a smaller area, but also has more symmetry attributed to the increased proportion of bargaining power of other agencies (e.g. RET). Due to the reduced service capacity, the second highest recipient of the cost allocation is shifted from HTM in the base setting (Table 11) to RET. This suggests that the reduced service capacity from NS is covered by RET, which lends the latter more negotiating power in setting up the agreement. In general, the capacity reduction of NS impacts the stability of allocations more and the total savings achieved through resource pooling less.

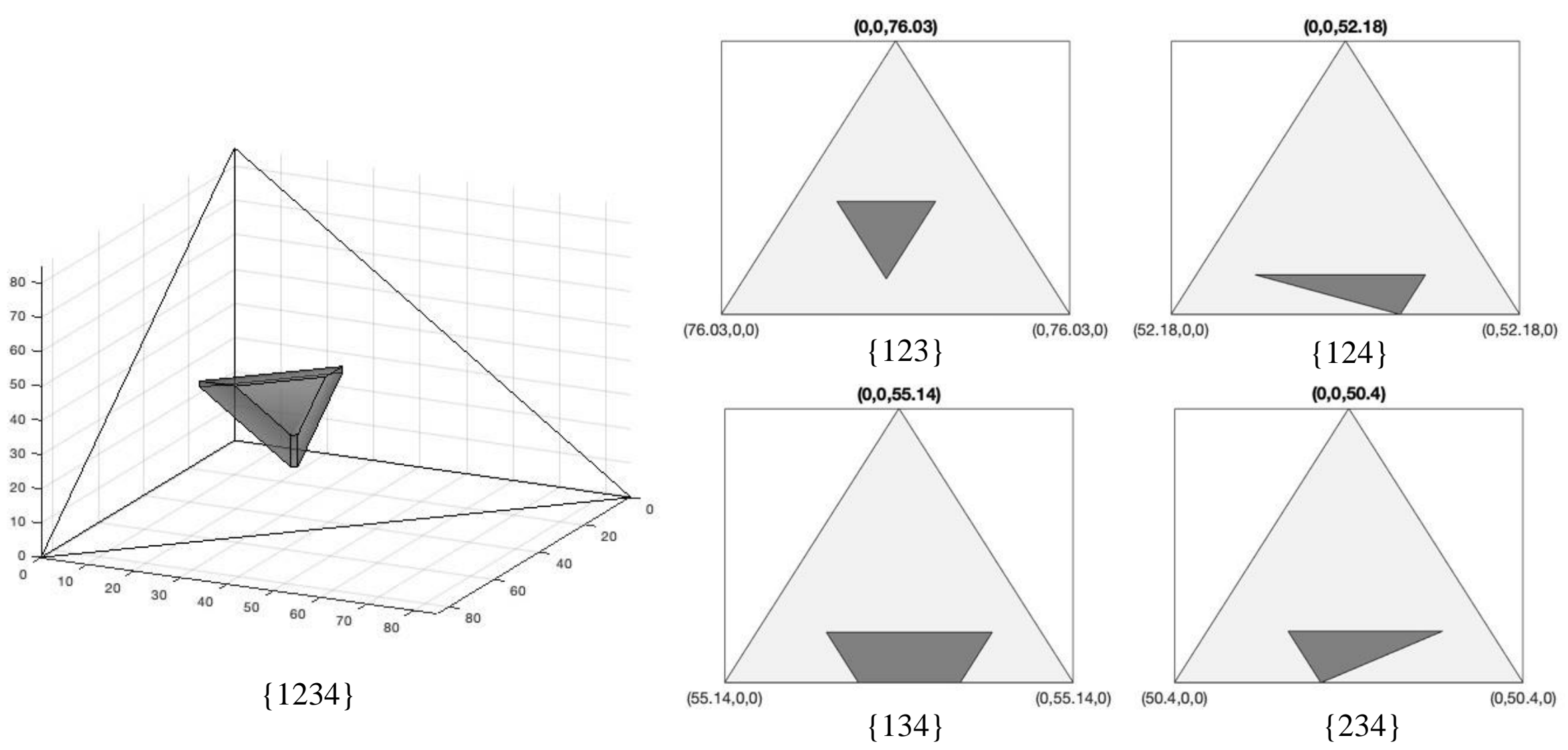

**Figure 10.** Core results for the Randstad operators under the 20% NS service capacity scenario.

## 6. Conclusion

With the emergence of a plethora of transportation services and the deregulation of traditional transport services, risk-pooling contracts will become vital in urban transit operations. New government policies should include subsidies as incentives for operators to share fleet resources. In this study, we propose a new mechanism for operators to set pre-disruption agreements to coordinate fleet resources during disruptions. This mechanism involves operators forming coalitions and committing fleet service capacity to a pool that can be accessed by any operators in the coalition during a disruption. A novel two-stage stochastic programming model is proposed to estimate transportation cost savings for a given coalition and classic cost allocation mechanisms are used to identify stable coalitions. The stochastic programming coalition valuation is solved using the L-shaped method along with SAA. Numerical experiments using realistic data from the Randstad public transport network in The Netherlands demonstrate the scalability of the method and its effectiveness in producing stable coalitions among four operators that can improve overall network performance by 66% during disruptions.

Furthermore, by relating the cost allocation mechanism to a stochastic programming model, we can evaluate the sensitivity of an operator's bargaining power to coordinate for disruptions to both their multicommodity flow network structure as well as to scenario distributions, which is a first in the literature. This is also demonstrated with the Randstad network, where the NS operator's bargaining power is shown to be sensitive to changes in network structure.

There are several directions for future research. The current resource sharing formulation does not integrate subsidies into the pool contribution decisions. However, we are also looking into a closed-loop side-payment mechanism that addresses stability as an equilibrium concept. In this case no external subsidies will be needed to ensure contract stability. Instead, a side-payment mechanism between operators would be introduced to maintain the social optimum solution. There are also many other fields that this work can be applied to beyond urban transportation: freight, airlines, other two-sided markets, and other network flow games where resource sharing is critical to face demand/supply uncertainty. Finally, the exchangeable units of link capacities introduced in this study as service capacity sharing have certain limitations that may be addressed in the context of asymmetric/one-way sharing where units are not completely interchangeble, but subject to physical limitations (e.g. buses may be able to substitute rail service but not vice versa). A subsequent study may investigate the benefits associated with having interoperable vehicular technologies due to the opportunity for risk pooling.

## Acknowledgements

The authors thank Menno Yap for processing and preparing the empirical data that used given as input to the case study analysis, and to Mengyun Mandy Li for preparing the transit network GIS figure. The first two authors were supported by NSF CMMI-1634973 and C2SMART University Transportation Center (USDOT #69A3551747124). The third author was supported by the CriticalMaaS project (no. 804469) which is financed by the European Research Council and the Amsterdam Institute of Advanced Metropolitan Solutions.

## References

Agarwal, R. and Ergun, Ö. (2008). Mechanism design for a multicommodity flow game in service network alliances. *Operations Research Letters*, 36(5), 520–524.

Asadabadi, A., & Miller-Hooks, E. (2020). Maritime port network resiliency and reliability through co-opetition. *Transportation Research Part E: Logistics and Transportation Review*, 137, 101916.

Banister, D. J., & Mackett, R. L. (1990). The minibus: theory and experience, and their implications. *Transport Reviews*, *10*(3), 189-214.

Barbarosoğlu, G. and Arda, Y. (2004). A two-stage stochastic programming framework for transportation planning in disaster response. *Journal of the Operational Research Society* 55(1), 43–53.

Bell, M. G., Kanturska, U., Schmöcker, J. D., & Fonzone, A. (2008). Attacker–defender models and road network vulnerability. *Philosophical Transactions of the Royal Society A: Mathematical, Physical and Engineering Sciences*, *366*(1872), 1893-1906.

Bird, C. G. (1976). On cost allocation for a spanning tree: a game theoretic approach. *Networks*, 6(4), 335-350.

Birge, J. R. and Louveaux, F. (2011). *Introduction to stochastic programming*. Springer.

Cats, O., Yap, M., and van Oort, N. (2016). Exposing the role of exposure: Public transport network risk analysis. *Transportation Research Part A* 88, 1–14.

Cats, O. and Jenelius, E. (2015). Planning for the unexpected: The value of reserve capacity for public transport network robustness. *Transportation Research Part A* 81, 47–61.

Cebecauer, M., Burghout, W., Jenelius, E., Babicheva, T., & Leffler, D. (2021). Integrating Demand Responsive Services into Public Transport Disruption Management. *IEEE Open Journal of Intelligent Transportation Systems*, 1-1.

Chen, A., & Yang, C. (2004). Stochastic transportation network design problem with spatial equity constraint. *Transportation Research Record*, 1882(1), 97-104.

Chen, A., Yang, H., Lo, H. K., & Tang, W. H. (1999). A capacity related reliability for transportation networks. *Journal of advanced transportation*, 33(2), 183-200.

Chow, J. Y. J., & Regan, A. C. (2014). A surrogate-based multiobjective metaheuristic and network degradation simulation model for robust toll pricing. *Optimization and Engineering*, 15(1), 137-165.

Chow, J. Y. J., & Sayarshad, H. R. (2014). Symbiotic network design strategies in the presence of coexisting transportation networks. *Transportation Research Part B: Methodological*, 62, 13-34.

Church, R. L., & Scaparra, M. P. (2007). Protecting critical assets: the r-interdiction median problem with fortification. *Geographical Analysis*, *39*(2), 129-146.

Consilvio, A., Di Febbraro, A., Moretti, V., & Sacco, N. (2020). Design of collaborative multimodal strategies for disruption management in urban railway systems. In *2020 IEEE 23rd International Conference on Intelligent Transportation Systems (ITSC)* (pp. 1-6).

Cruijssen, F., Bräysy, O., Dullaert, W., Fleuren, H., & Salomon, M. (2007). Joint route planning under varying market conditions. *International Journal of Physical Distribution & Logistics Management* 37(4), 287-304.

Curiel, I., Derks, J., & Tijs, S. (1989). On balanced games and games with committee control. *Operations-Research-Spektrum*, 11(2), 83-88.

D'Amours, S., & Rönnqvist, M. (2010). Issues in collaborative logistics. In *Energy, natural resources and environmental economics* (pp. 395-409). Springer, Berlin, Heidelberg.

Derks, J. J. M., & Tijs, S. H. (1985). *Stable outcomes for multi-commodity flow games*. Tilburg University, School of Economics and Management.

Doukidis, G. I., Mason, R., Lalwani, C., & Boughton, R. (2007). Combining vertical and horizontal collaboration for transport optimisation. *Supply Chain Management: An International Journal* 12(3), 187-199.

Drechsel, J., & Kimms, A. (2010). Computing core allocations in cooperative games with an application to cooperative procurement. *International Journal of Production Economics*, 128(1), 310-321.

Drechsel, J., & Kimms, A. (2011). Cooperative lot sizing with transshipments and scarce capacities: solutions and fair cost allocations. *International Journal of Production Research*, 49(9), 2643-2668.

Driessen, T. (1988). Cooperative Games and Examples. In *Cooperative Games, Solutions and Applications* (pp. 1-12). Springer, Dordrecht.

Elçi, Ö., & Noyan, N. (2018). A chance-constrained two-stage stochastic programming model for humanitarian relief network design. *Transportation research part B* 108, 55-83.

Fang, Y., Jiang, Y., & Fei, W. (2020). Disruption Recovery for Urban Public Tram System: An Analysis of Replacement Service Selection. *IEEE Access*, 8, 31633-31646.

Fotouhi, H., Moryadee, S., & Miller-Hooks, E. (2017). Quantifying the resilience of an urban traffic-electric power coupled system. *Reliability Engineering & System Safety*, 163, 79-94.

Frasinelli, M. (2012). NJ Transit head puts Sandy damage estimate at $400M. *The Star Ledger*, Dec 6.

Gilles, D. (1953). *Some theorems on n-person games*. Ph.D. Dissertation, Princeton University, Department of Mathematics.

González-Díaz, J., & Sánchez-Rodríguez, E. (2007). A natural selection from the core of a TU game: the core-center. *International Journal of Game Theory*, 36(1), 27-46.

Grass, E., & Fischer, K. (2016). Two-stage stochastic programming in disaster management: A literature survey. *Surveys in Operations Research and Management Science*, 21(2), 85-100.

Gu, W., Yu, J., Ji, Y., Zheng, Y., & Zhang, H. M. (2018). Plan-based flexible bus bridging operation strategy. *Transportation Research Part C: Emerging Technologies*, 91, 209-229.

Haghani, A., & Oh, S. C. (1996). Formulation and solution of a multi-commodity, multi-modal network flow model for disaster relief operations. *Transportation Research Part A* 30(3),

231-250.
Hensher, D. A. (2017). Future bus transport contracts under a mobility as a service (MaaS) regime in the digital age: Are they likely to change? *Transportation Research Part A: Policy and Practice*, 98, 86-96.
Hong, X., Lejeune, M. A., & Noyan, N. (2015). Stochastic network design for disaster preparedness. *IIE Transactions*, 47(4), 329-357.
Hoppe, R. H. W. (2007) *Chapter 1: Stochastic Linear and Nonlinear Programming*. Lecture notes, https://www.semanticscholar.org/paper/Chapter-1-Stochastic-Linear-and-Nonlinear-Hoppe/ef6ce5358781eb2ffa6b2a44e87edde000c1948f.
Hu, H., Gao, Y., Yu, J., Liu, Z., & Li, X. (2016). Planning bus bridging evacuation during rail transit operation disruption. *Journal of Urban Planning and Development*, 142(4), 04016015.
Jin, J. G., Lu, L., Sun, L., & Yin, J. (2015). Optimal allocation of protective resources in urban rail transit networks against intentional attacks. *Transportation Research Part E: Logistics and Transportation Review*, 84, 73-87.
Jin, J. G., Teo, K. M., & Odoni, A. R. (2016). Optimizing bus bridging services in response to disruptions of urban transit rail networks. *Transportation Science*, 50(3), 790-804.
Kalai, E., & Zemel, E. (1982). Generalized network problems yielding totally balanced games. *Operations Research*, 30(5), 998-1008.
Kellner, F., & Otto, A. (2012). Allocating CO 2 emissions to shipments in road freight transportation. *Journal of Management Control*, 22(4), 451-479.
Kepaptsoglou, K., & Karlaftis, M. G. (2009). The bus bridging problem in metro operations: conceptual framework, models and algorithms. Public Transport, 1(4), 275-297.
Klibi, W., Ichoua, S., & Martel, A. (2018). Prepositioning emergency supplies to support disaster relief: A case study using stochastic programming. *INFOR 56*(1), 50-81.
Kolker, A. (2018). The concept of the Shapley value and the cost allocation between cooperating participants. In *Encyclopedia of Information Science and Technology, Fourth Edition* (pp. 2095-2107). IGI Global.
Levi, D.S., Kaminsky, P. and Simchi-Levi, E., 2003. Chapter 3: Inventory Management and Risk Pooling. *Designing & Managing the Supply Chain, Second Edition* (p-66).
Li, J., & Wang, X. (2020). Multimodal evacuation after subway breakdown: a modeling framework and mode choice behavior. *Transportation research interdisciplinary perspectives*, 6, 100177.
Liu, C., Fan, Y., & Ordóñez, F. (2009). A two-stage stochastic programming model for transportation network protection. *Computers & Operations Research*, 36(5), 1582-1590.
Liu, Q. (2022). Data use and sharing in public transit systems. PhD dissertation, New York University, 166p.
Liu, Q., & Chow, J. Y. J. (2022). Efficient and stable data-sharing in a public transit oligopoly as a coopetitive game. *Transportation Research Part B: Methodological*, 163, 64-87.
Lo, H. K., & Tung, Y. K. (2003). Network with degradable links: capacity analysis and design. *Transportation Research Part B: Methodological*, 37(4), 345-363.
Lozano, S., Moreno, P., Adenso-Díaz, B., & Algaba, E. (2013). Cooperative game theory approach to allocating benefits of horizontal cooperation. *European Journal of Operational Research*, 229(2), 444-452.
Ma, T. Y., Rasulkhani, S., Chow, J. Y., & Klein, S. (2019). A dynamic ridesharing dispatch and idle vehicle repositioning strategy with integrated transit transfers. *Transportation Research*

*Part E: Logistics and Transportation Review*, 128, 417-442.
Megiddo, N. (1978). Cost allocation for Steiner trees. *Networks*, 8(1), 1-6.
Meinhardt, H. I., (2020). MatTuGames: A Matlab Toolbox for Cooperative Game Theory. https://www.mathworks.com/matlabcentral/fileexchange/35933-mattugames. (Accessed 06.10.20).
Miandoabchi, E., & Farahani, R. Z. (2011). Optimizing reserve capacity of urban road networks in a discrete network design problem. *Advances in Engineering Software*, 42(12), 1041-1050.
Miller-Hooks, E., Zhang, X., & Faturechi, R. (2012). Measuring and maximizing resilience of freight transportation networks. *Computers & Operations Research*, 39(7), 1633-1643.
Mirás Calvo, M.A., Sánchez Rodriguez, E., (2006). TUGlab: A Cooperative Game Theory Toolbox. http://mmiras.webs.uvigo.es/TUGlab/. (Accessed 06.10.20).
Myerson, R. B. (1980). Conference structures and fair allocation rules. *International Journal of Game Theory*, 9(3), 169-182.
Neves-Silva, R. (2013). Coping with disruptions in public tram systems: cases in Germany, China and the United States. In *Intelligent Decision Technologies: Proceedings of the 5th KES International Conference on Intelligent Decision Technologies (KES-IDT 2013)* (Vol. 255, p. 150). Courier Corporation.
Noyan, N. (2012). Risk-averse two-stage stochastic programming with an application to disaster management. *Computers & Operations Research*, 39(3), 541-559.
Owen, G. (1982). *Game theory*. Academic Press, New York.
Özener, O. Ö., & Ergun, Ö. (2008). Allocating costs in a collaborative transportation procurement network. *Transportation Science*, 42(2), 146-165.
Pantelidis, T., L-Shaped_runtimes, (2020), GitHub repository, https://github.com/theodoros1993/L-Shaped_runtimes. (Accessed 06.10.20).
Pantelidis, T. P., Chow, J. Y., & Rasulkhani, S. (2020). A many-to-many assignment game and stable outcome algorithm to evaluate collaborative mobility-as-a-service platforms. *Transportation Research Part B: Methodological*, 140, 79-100.
Peeta, S., Salman, F. S., Gunnec, D., & Viswanath, K. (2010). Pre-disaster investment decisions for strengthening a highway network. *Computers & Operations Research*, 37(10), 1708-1719.
Pender, B., Currie, G., Delbosc, A., & Shiwakoti, N. (2013). Disruption recovery in passenger railways: International survey. *Transportation research record*, 2353(1), 22-32.
Rasulkhani, S., & Chow, J. Y. J. (2019). Route-cost-assignment with joint user and operator behavior as a many-to-one stable matching assignment game. *Transportation Research Part B: Methodological, 124,* 60-81.
Rawls, C. G., & Turnquist, M. A. (2010). Pre-positioning of emergency supplies for disaster response. *Transportation research part B: Methodological*, 44(4), 521-534.
Reinhardt, G., & Dada, M. (2005). Allocating the gains from resource pooling with the Shapley value. *Journal of the Operational Research Society*, 56(8), 997-1000.
Saberi, M., Ghamami, M., Gu, Y., Shojaei, M. H. S., & Fishman, E. (2018). Understanding the impacts of a public transit disruption on bicycle sharing mobility patterns: A case of Tube strike in London. *Journal of Transport Geography*, 66, 154-166.
Schmeidler, D. (1969). The nucleolus of a characteristic function game. *SIAM Journal on applied mathematics*, 17(6), 1163-1170.
Schotanus, F., Telgen, J., & de Boer, L. (2008). Unfair allocation of gains under the Equal Price

allocation method in purchasing groups. *European Journal of Operational Research*, *187*(1), 162-176.

Shapiro, A., & Philpott, A. (2007). *A tutorial on stochastic programming*. Lecture notes, https://stoprog.org/sites/default/files/SPTutorial/TutorialSP.pdf

Shapley, L. S. (1951). *Notes on the n-Person Game — II: The Value of an n-Person Game*. RAND Corporation.

Smith, J. C., & Lim, C. (2008). Algorithms for network interdiction and fortification games. In *Pareto optimality, game theory and equilibria* (pp. 609-644). Springer, New York, NY.

Sumalee, A., & Watling, D. P. (2008). Partition-based algorithm for estimating transportation network reliability with dependent link failures. *Journal of Advanced Transportation*, 42(3), 213-238.

Tijs, S. H., & Driessen, T. S. (1986). Game theory and cost allocation problems. *Management science*, *32*(8), 1015-1028.

Tyndall, J. (2019). Free-floating carsharing and extemporaneous public transit substitution. *Research in Transportation Economics*, 74, 21-27.

Tuljak-Suban, D. (2018). Game Theory - Applications in Logistics and Economy. *Game Theory - Applications in Logistics and Economy.* InTech.

Van der Hurk, E., Koutsopoulos, H. N., Wilson, N., Kroon, L. G., & Maróti, G. (2016). Shuttle planning for link closures in urban public transport networks. *Transportation Science*, 50(3), 947-965.

Van Slyke, R. M., & Wets, R. (1969). L-shaped linear programs with applications to optimal control and stochastic programming. *SIAM Journal on Applied Mathematics*, 17(4), 638-663.

Wang, H., Lam, W. H., Zhang, X., & Shao, H. (2015). Sustainable transportation network design with stochastic demands and chance constraints. *International Journal of Sustainable Transportation*, 9(2), 126-144.

Yang, Z., & Chen, X. (2019). Compensation decisions on disruption recovery service in urban rail transit. *Promet-Traffic&Transportation*, 31(4), 367-375.

Yoon, G., Chow, J. Y., & Rath, S. (2022). A simulation sandbox to compare fixed-route, semi-flexible transit, and on-demand microtransit system designs. *KSCE Journal of Civil Engineering*, *26*(7), 3043-3062.

Zeng, A. Z., Durach, C. F., & Fang, Y. (2012). Collaboration decisions on disruption recovery service in urban public tram systems. *Transportation research part E: logistics and transportation review*, *48*(3), 578-590.

Zhang, S., & Lo, H. K. (2018). Metro disruption management: Optimal initiation time of substitute bus services under uncertain system recovery time. *Transportation Research Part C: Emerging Technologies*, 97, 409-427.

Zhang, S., & Lo, H. K. (2020). Metro disruption management: Contracting substitute bus service under uncertain system recovery time. *Transportation Research Part C: Emerging Technologies*, 110, 98-122.

Zhang, T., Niu, C., Jayakumar Nair, D., Robson, E. N., & Dixit, V. (2021). Transportation resilience optimization from an economic perspective at the pre-event stage. *Available at SSRN 4099925*.